\documentclass[twocolumn]{aastex63}

\def\a{\alpha}
\def\b{\beta}
\def\g{\gamma}
\def\y{\gamma}
\def\d{\delta}
\def\D{\Delta}
\def\e{\epsilon}
\def\i{\iota}
\def\k{\kappa}
\def\l{\lambda}
\def\u{\mu}
\def\v{\nu}
\def\s{\sigma}
\def\t{\tau}
\def\th{\theta}
\def\x{\xi}

\def\n{\nabla}
\def\w{\omega}
\def\z{\zeta}

\def\p{\phi}

\def\L{\Lambda}

\def\O{\Omega}

\shorttitle{\textsc{Two Polars from ZTF/eFEDS}}
\shortauthors{Rodriguez et al.}
\graphicspath{{./}{figures/}}

\usepackage{float}
\usepackage{appendix}
\usepackage{xcolor}
\usepackage{amsmath}

\begin{document}

\title{Discovery of Two Polars from a Crossmatch of ZTF and the SRG/eFEDS X-ray Catalog} 

\correspondingauthor{Antonio C. Rodriguez}
\email{acrodrig@caltech.edu}

\author[0000-0003-4189-9668]{Antonio C. Rodriguez}

\affiliation{Department of Astronomy, California Institute of Technology, 
1200 E. California Blvd, Pasadena, CA, 91125, USA}

\author[0000-0001-5390-8563]{Shrinivas R. Kulkarni}
\affiliation{Department of Astronomy, California Institute of Technology, 
1200 E. California Blvd, Pasadena, CA, 91125, USA}

\author{Thomas A. Prince}
\affiliation{Division of Physics, Mathematics, and Astronomy, California Institute of Technology, Pasadena, CA 91125, USA}

\author[0000-0003-4373-7777]{Paula Szkody}
\affiliation{Department of Astronomy, University of Washington, 3910 15th Avenue NE, Seattle, WA 98195, USA}

\author[0000-0002-7226-836X]{Kevin B. Burdge}
\affiliation{MIT-Kavli Institute for Astrophysics and Space Research 77 Massachusetts Ave. Cambridge, MA 02139, USA}

\author[0000-0002-4770-5388]{Ilaria Caiazzo}
\affiliation{Department of Astronomy, California Institute of Technology, 
1200 E. California Blvd, Pasadena, CA, 91125, USA}

\author[0000-0002-2626-2872]{Jan van Roestel}
\affiliation{Department of Astronomy, California Institute of Technology, 
1200 E. California Blvd, Pasadena, CA, 91125, USA}

\author[0000-0002-0853-3464]{Zachary P. Vanderbosch}
\affiliation{Department of Astronomy, California Institute of Technology, 
1200 E. California Blvd, Pasadena, CA, 91125, USA}

\author[0000-0002-6871-1752]{Kareem El-Badry}
\affiliation{Center for Astrophysics $|$ Harvard \& Smithsonian, 60 Garden Street, Cambridge, MA 02138, USA}

\author[0000-0001-8018-5348]{Eric C. Bellm}
\affiliation{DIRAC Institute, Department of Astronomy, University of Washington, 3910 15th Avenue NE, Seattle, WA 98195, USA}

\author[0000-0002-2761-3005]{Boris T. Gänsicke}
\affiliation{Department of Physics, University of Warwick, Coventry CV4 7AL, UK}

\author[0000-0002-3168-0139]{Matthew J. Graham}
\affiliation{Department of Astronomy, California Institute of Technology, 
1200 E. California Blvd, Pasadena, CA, 91125, USA}

\author[0000-0003-2242-0244]{Ashish~A.~Mahabal}
\affiliation{Division of Physics, Mathematics and Astronomy, California Institute of Technology, Pasadena, CA 91125, USA}
\affiliation{Center for Data Driven Discovery, California Institute of Technology, Pasadena, CA 91125, USA}

\author[0000-0002-8532-9395]{Frank J. Masci}
\affiliation{IPAC, California Institute of Technology, 1200 E. California
             Blvd, Pasadena, CA 91125, USA}
             
\author[0000-0001-7016-1692]{Przemek Mróz}
\affiliation{Astronomical Observatory, University of Warsaw, 
Al. Ujazdowskie 4, 00-478, 
Warszawa, Poland}

\author[0000-0002-0387-370X]{Reed Riddle}
\affiliation{California Institute of Technology, Department of Astronomy, 
1200 E. California Blvd, Pasadena, CA, 91125, USA}

\author[0000-0001-7648-4142]{Ben Rusholme}
\affiliation{IPAC, California Institute of Technology, 1200 E. California
             Blvd, Pasadena, CA 91125, USA}

\begin{abstract}

Magnetic CVs are luminous Galactic X-ray sources but have been  difficult to find in purely optical surveys due to their lack of outburst behavior. The eROSITA telescope aboard the Spektr-RG (SRG) mission is conducting an all-sky X-ray survey and recently released the public eROSITA Final Equatorial Depth Survey (eFEDS) catalog. We crossmatched the eFEDS catalog with photometry from the Zwicky Transient Facility (ZTF) and discovered two new magnetic cataclysmic variables (CVs). We obtained high-cadence optical photometry and phase-resolved spectroscopy for each magnetic CV candidate and found them both to be polars. Among the newly discovered magnetic CVs is ZTFJ0850+0443, an eclipsing polar with orbital period $P_\textrm{orb} = 1.72$ hr, white dwarf mass $M_\textrm{WD} = 0.81 \pm 0.08 M_\odot$ and accretion rate $\dot{M} \sim 10^{-11} M_\odot$/yr. We suggest that ZTFJ0850+0443 is a low magnetic field strength polar, with $B_\textrm{WD} \lesssim 10$ MG. We also discovered a non-eclipsing polar, ZTFJ0926+0105, with orbital period $P_\textrm{orb} = 1.48$ hr, magnetic field strength $B_\textrm{WD} \gtrsim 26$ MG, and accretion rate $\dot{M} \sim 10^{-12} M_\odot$/yr. 
\end{abstract}

\section{Introduction}

Magnetic cataclysmic variables (CVs) are compact object binaries in which a highly magnetized white dwarf (WD) accretes from a Roche-lobe filling donor, typically a late-type main-sequence star. Magnetic CVs are interesting for two reasons: 1) they may be the dominant contributors to the Galactic ridge hard X-ray emission \citep{2016hailey} and 2) the origin of the strong ($B \sim $ 1--100 MG) magnetic fields in accreting WDs is uncertain \citep{2000oldmagrev, schreiber2021}. More broadly, they are rich laboratories for studying accretion under the influence of a strong magnetic field.

CVs typically consist of a WD accreting from a donor via an accretion disk \citep[e.g.][]{warner95, hellierbook}. In non-magnetic CVs, the secondary fills its Roche lobe and develops a teardrop-like shape with the tip positioned at the Lagrangian L1 point. Matter leaves the secondary star through this point and forms an accretion stream after exiting the donor star. This stream extends out until a point known as the ``circularization radius". The circularization radius is the point where the matter in the accretion stream intersects itself as it orbits the primary. The material then forms an accretion disk around the white dwarf and makes its way to the surface through viscous dissipation of energy. 

In magnetic CVs known as intermediate polars (IPs; $B_\textrm{WD} \approx  1-10$ MG), the magnetic field pressure is comparable to the ram pressure of the accreted material. As a result, the disk is truncated at the Alfvén radius (also known as the magnetospheric radius). In IPs, the Alfvén radius is smaller than the circularization radius so that a partial disk forms. The accreted material initially flows through the disk but is channeled along field lines onto the white dwarf surface inside the Alfvén radius. 

In magnetic CVs known as polars ($B_\textrm{WD} \gtrsim  10$ MG), the magnetic field is strong enough that the Alfvén radius is larger than the circularization radius \citep[e.g.][]{hellierbook, 2017mukai}. Matter can build up at the Alfvén radius, in a region called the stagnation region (also referred to as the threading region), before being channeled by WD magnetic field lines onto the surface \citep[e.g.][]{1990cropper, 2017mukai}.

Extensive literature and evolutionary models exist for non-magnetic CVs \citep[e.g.][]{2011knigge}.
In the canonical picture of non-magnetic CVs, these systems are formed through common envelope evolution. Angular momentum loss (AML) of the binary system then drives its evolution as the white dwarf accretes matter from its secondary companion. At orbital periods above $\approx$ 3 hrs, magnetic braking dominates over gravitational radiation as the dominant contributor to AML. Magnetic braking is thought to shut off at the point when the donor star becomes fully convective, leading to the observed period gap ($P_\textrm{orb}\approx$ 2--3 hrs). Below this gap, gravitational radiation dominates the loss of angular momentum. Few CVs have been found at orbital periods below $\approx$ 80 mins (known as the "period bounce"), thought to correspond to the point where the donor star becomes degenerate and expands as it loses mass.

There is typically no mention of WD magnetic fields in this evolutionary model of CVs. The discovery of many magnetic CVs within the period gap, and the large number of polars, at periods below 2 hours, has led to new ideas on how magnetic CVs form and evolve \citep[e.g.][]{2020belloni, schreiber2021}.

Magnetic CVs have previously been difficult to find via optical surveys alone. Both magnetic and non-magnetic CVs have historically been discovered through their novae or dwarf novae outburst behavior. The former occurs when a thermonuclear ignition of hydrogen occurs on or near the surface of the accreting white dwarf. The latter occurs when a thermal instability in the accretion disk leads to a temporary increase in accretion rate \citep{hellierbook}. However, searching via optical outbursts alone is inefficient for finding magnetic CVs. The magnetic field disrupts the disk or eliminates it entirely, which prevents a thermal instability \citep{2017hameury_dn_ip}. Novae can still occur in magnetic CVs, but are rare. As a result, optical-only surveys lead to a $\sim 1/100$ rate of discovery for magnetic CVs \citep[e.g.][]{ztf_cv_yr2}. However, a recent volume-limited ($d < 150$ pc) study of CVs discovered via various techniques found 36\% of CVs to be magnetic \citep{pala2020}.

Both magnetic and non-magnetic CVs are strong X-ray emitters. The source of X-rays in magnetic CVs is typically thermal bremsstrahlung from the shock of material accreting onto the WD surface \citep[e.g.][]{1990cropper, 2000oldmagrev, 2017mukai}. The resulting thermal bremsstrahlung temperature is calculated via the following equation:
\begin{gather}
    kT_\textrm{shock, bremss} \approx \frac{3}{8}\frac{GM_\textrm{WD}\mu m_H}{R_\textrm{WD}}
\end{gather}
which for a white dwarf of mass $0.8 M_\odot$ is 36 keV --- well in the regime of hard X-rays. 

The source of X-rays in non-magnetic CVs can be difficult to disentangle: the hot WD photosphere of a recent nova, the optically thick boundary layer, or accretion disk wind shock of non-magnetic CVs can also lead to X-ray emission \citep{2017mukai}. 

X-ray surveys help overcome the observational bias of optical-only surveys in finding CVs by uncovering both the magnetic and non-magnetic populations \citep[e.g.][]{1996motch, 2017bernardini, 2018halpern}. Many magnetic CVs were discovered through the transformative all-sky Roentgensatellit X-ray survey \citep[ROSAT;][]{1982rosat1, 1999rosat2, 2016rosat3}. The ongoing eROSITA telescope aboard the Spektr-RG mission \citep[SRG;][]{2021sunyaev, 2021erosita} is projected to go $\sim$30 times deeper than ROSAT with improved localization ($\sim$5 arcsec) of X-ray sources, guaranteeing discoveries of new magnetic CVs, among other objects, all over the sky.

In this work, we crossmatched the eROSITA Final Equatorial Depth Survey \citep[eFEDS;][]{salvato2021} catalog with forced photometry of ZTF Data Release 5 (DR5)\footnote{\url{https://www.ztf.caltech.edu/ztf-public-releases.html}}.  We discovered two new polars: ZTFJ0850+0443 and ZTFJ0926+0105. In Section \ref{sec:catalog}, we present an overview of the SRG/eFEDS catalog and ZTF archival photometry. In Section \ref{sec:methodology}, we outline our methodology for finding notable objects within the crossmatched dataset. In Section \ref{sec:0850}, we present follow-up high-cadence photometry and spectroscopy and analyze all data for ZTFJ0850+0443. We present data and analysis for ZTFJ0926+0105 in Section \ref{sec:0926}. Finally, in Section \ref{sec:discussion}, we place this work in the context of previous CV studies and discuss the utility of X-ray/optical searches for finding otherwise elusive CVs.

\section{Catalog Data}
\label{sec:catalog}
\subsection{eFEDS Catalog}
We began with the catalog of eFEDS Galactic sources. \cite{salvato2021} searched for optical counterparts to the eFEDS X-ray sources by crossmatching to the DECam Legacy Survey (DECaLS) LS8 catalog, part of the DESI Legacy Imaging Survey \citep{2019desi_surveys}. The median separation between the X-ray source and optical counterpart, divided by the mean X-ray positional error (i.e. the eROSITA/SRG uncertainty in the localization of the X-ray source), is reported in \cite{salvato2021} to be 1.22. For the crossmatch to ZTF data, we use the coordinates of the LS8 optical counterpart. Given the superior depth of DECaLS (23.4 mag), we crossmatch the sources to the LS8 catalog instead of the ZTF catalog (21 mag).

Galactic sources in eFEDS are mainly identified through two methods as described in \cite{salvato2021}: 1) SDSS redshifts being $z < 0.002$ and/or 2) \textit{Gaia} parallaxes with good significance: $\pi/\sigma_\pi > 3$. The Renormalised Unit Weight Error (RUWE) $<$ 1.4 can also be used. The actual method described in \cite{salvato2021} is complex, and uses a combination of NWAY, a tool based on Bayesian statistics, and ASTROMATCH, a tool based on the Maximum Likelihood Ratio. 

In the eFEDS Main Catalog, 24774/27369 (90.5\%) of X-ray sources are reported to have a reliable optical counterpart. Of those sources, 2976 are classified with the label \texttt{LIKELY GALACTIC} or \texttt{SECURE GALACTIC}. It is this sample that we further investigate with ZTF.

\subsection{ZTF Data}
The Zwicky Transient Facility (ZTF) is a photometric survey that uses a wide 47 $\textrm{deg}^2$ field-of-view camera mounted on the Samuel Oschin 48-inch telescope at Palomar Observatory with $g$, $r$, and $i$ filters \citep{bellm2019, graham2019, dekanyztf, masci_ztf}. In its first year of operations, ZTF carried out a public nightly Galactic Plane Survey in $g$-band and $r$-band \citep{ztf_northernskysurvey_bellm, kupfer_ztf}. This survey was in addition to the Northern Sky Survey which operated on a 3 day cadence \citep{bellm2019}. Since entering Phase II, the public Northern Sky Survey is now at a 2-day cadence. The pixel size of the ZTF camera is 1" and the median delivered image quality is 2.0" at FWHM. 

We use forced photometry from ZTF Data Release 5 (DR5). Lightcurves have a photometric precision of 0.01 mag at 13--14 mag down to a precision of 0.1--0.2 mag for the faintest objects at 20--21 mag. While both raw photometry and forced photometry are PSF-fit photometry, the forced photometry calculates photometry of the object on difference images by forcing the location of the PSF to remain fixed according to the ZTF absolute astrometric reference. This allows one to obtain flux estimates below the detection threshold and therefore probe deeper than the standard photometry. For bulk download, we use a database of forced photometry files (Mróz, Burdge, et al. in prep).

\subsection{Objects of Interest}

\begin{figure}
    \centering
    \includegraphics[scale=0.34]{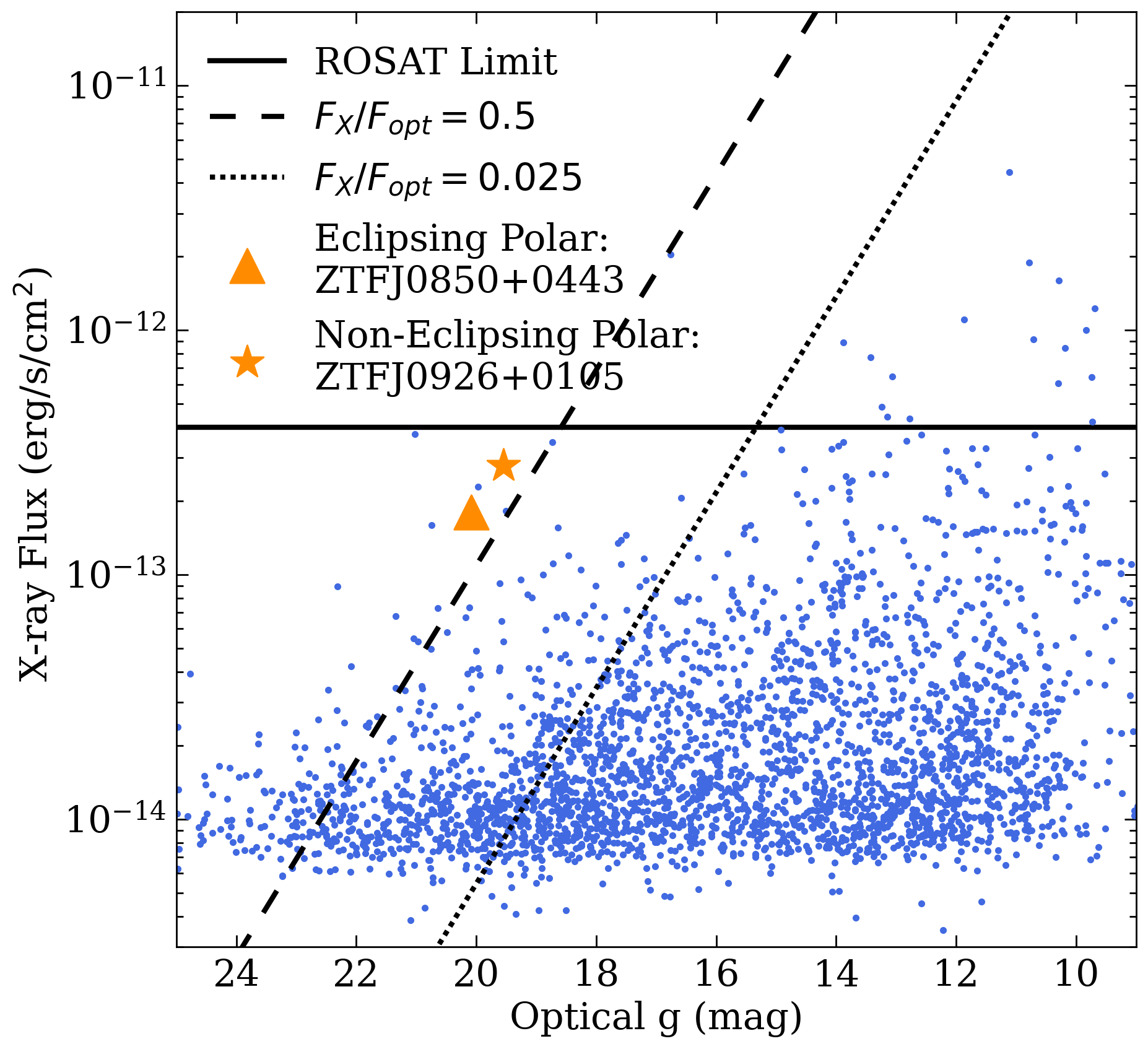}
    \caption{X-ray flux compared to DECaLS LS8 optical flux with lines of constant $F_X/F_\textrm{opt}$ shown. All SRG/eFEDS Galactic objects are shown in blue, virtually all of which were below the detection threshold of the ROSAT all-sky X-ray survey. The two polars stand out as systems with some of the largest $F_X/F_\textrm{opt}$ ratios in the catalog.}
    \label{fig:xray_opt}
\end{figure}

It is useful to plot the X-ray to optical flux ratio for the entire eFEDS catalog. A similar exercise was carried out by various studies \citep[e.g.][]{2009agueros, 2015greiner} for the ROSAT catalog and proved to be an effective way for disentangling CVs from other Galactic X-ray sources such as M dwarfs and chromospherically active binaries such as BY Dra and RS CVn systems. Figure \ref{fig:xray_opt} shows the eFEDS X-ray flux versus DECaLS LS8 optical flux plot along with ZTFJ0850+0443 and ZTFJ0926+0105. Both systems stand out as being amongst the systems with the highest $F_X/F_\textrm{opt}$ ratio.  

\begin{figure}
    \centering
    \includegraphics[scale=0.34]{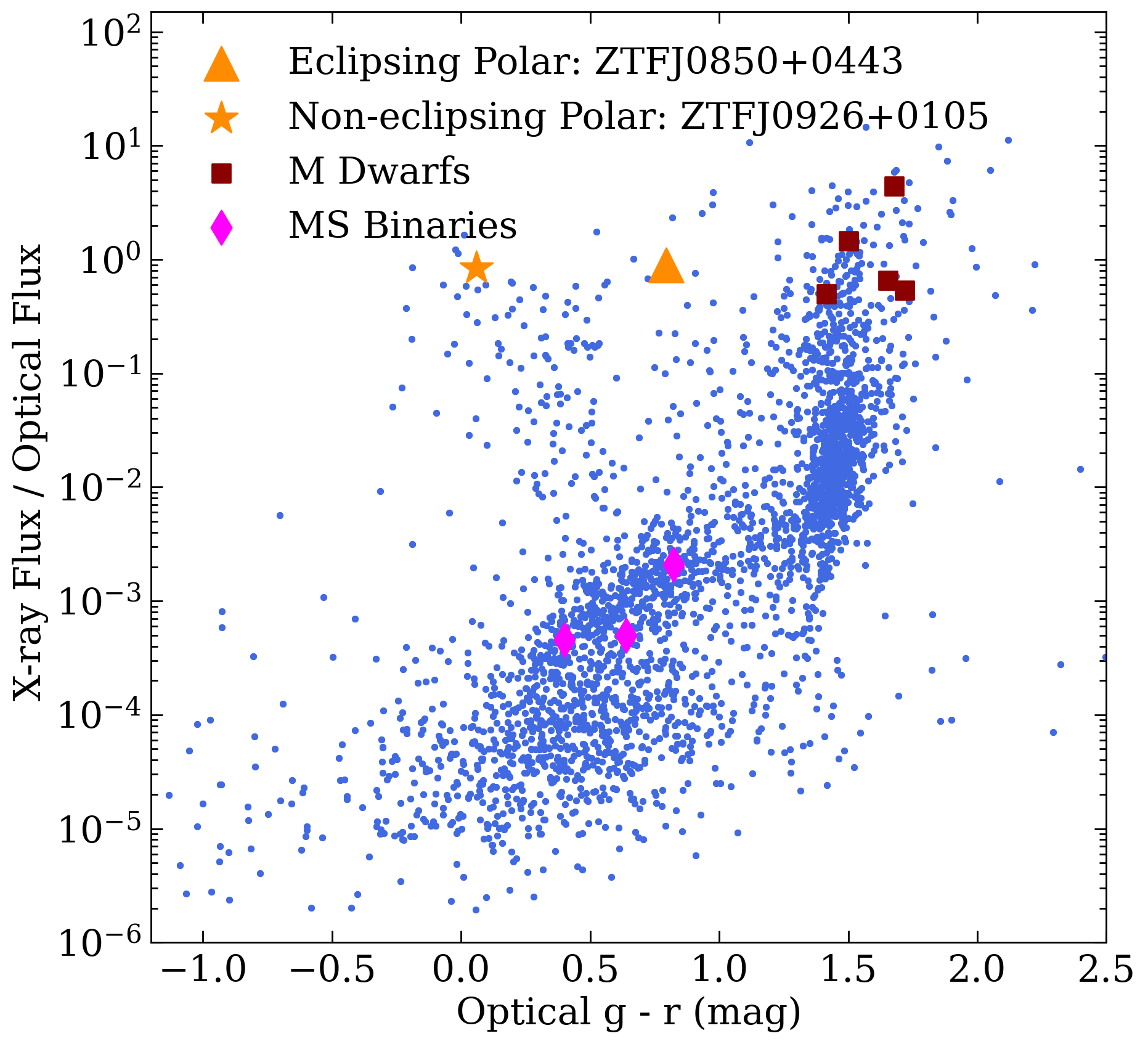}
    \caption{Ratio of X-ray flux to DECaLS LS8 optical flux as a function of optical color. The two polars cluster towards the upper left. Active M dwarfs dwarfs cluster towards the upper right and active main sequence binaries (known as BY Dra or RS CVn) below.}
    \label{fig:alternate}
\end{figure}

Figure \ref{fig:alternate} shows the advantage of a color cut in picking out CVs from active M dwarfs. By supplementing the X-ray to optical flux ratio with optical color information, M dwarfs cluster towards the upper right, while CVs cluster towards the upper left. The population in the lower center of the figure is generally comprised of the chromsopherically and coronally active BY Dra/RS CVn systems.\footnote{The term BY Dra refers to chromospherically active binary stars on the main sequence, and should be distinguished from RS CVn binary stars which are evolved and therefore above the main sequence. However, the term RS CVn has been used in the literature to refer to both; see \cite{1992bydra_rscvn} for a summary of the nomenclature.}

\section{Sample Selection}
\label{sec:methodology}
The goal of our study is to classify Galactic eFEDS/ZTF sources, with an emphasis on those with a high X-ray to optical flux ratio and a strong periodic signal in ZTF. 

We adopt the value of optical flux, $F_\textrm{opt}$, for our entire study, as the standard conversion of the DECaLS LS8 $g$ magnitude from the AB magnitude system. We assume a flat SED and a reference wavelength of 5000 Angstrom. We define the following samples:
 
 \begin{enumerate}
     \item High X-ray to optical flux ratio. We obtain spectra for all objects with $F_X/F_\textrm{opt} > 0.5$. 
     \item Moderate X-ray to optical flux ratio and strong periodicity. We obtain spectra for all objects with $0.5> F_X/F_\textrm{opt} > 0.025$ percent that feature strong periodicity and pass a color cut. Our color cut eliminates the reddest objects which are likely to be active M dwarfs. We also report best-fit periods for the objects that do not pass the color cut. 
     \item Low X-ray to optical flux ratio and strong periodicity. We compile a list periods for all objects that have $F_X/F_\textrm{opt} < 0.025$.
 \end{enumerate}

We define the significance of periodicity as the maximum Lomb-Scargle power subtracted by the median power, all divided by the median absolute deviation. We define ``strong periodicity" as lightcurves where the significance is in the $86^\textrm{th}$ quantile. By this metric, approximately 10 percent of all objects in the eFEDS/ZTF footprint show strong periodicity. We define our color cut in the LS8 color bands and exclude objects with $g - r > 1.4$ as likely M dwarfs.
 
ZTFJ0850+0443 and ZTFJ0926+0105 stood out immediately from the first cut, with values of $F_X/F_\textrm{opt} > 0.5$. The Lomb-Scargle periodogram search of their lightcurves using \texttt{gatspy} also showed strong periodicity \citep{gatspy, 2018vanderplas}. Both criteria prompted follow-up spectroscopy. 

The full analysis of our findings from the above cuts will be reported in an upcoming study. 

\section{ZTFJ0850+0443}
\label{sec:0850}

\subsection{Data}
\begin{figure}
    \centering
    \includegraphics[scale=0.13]{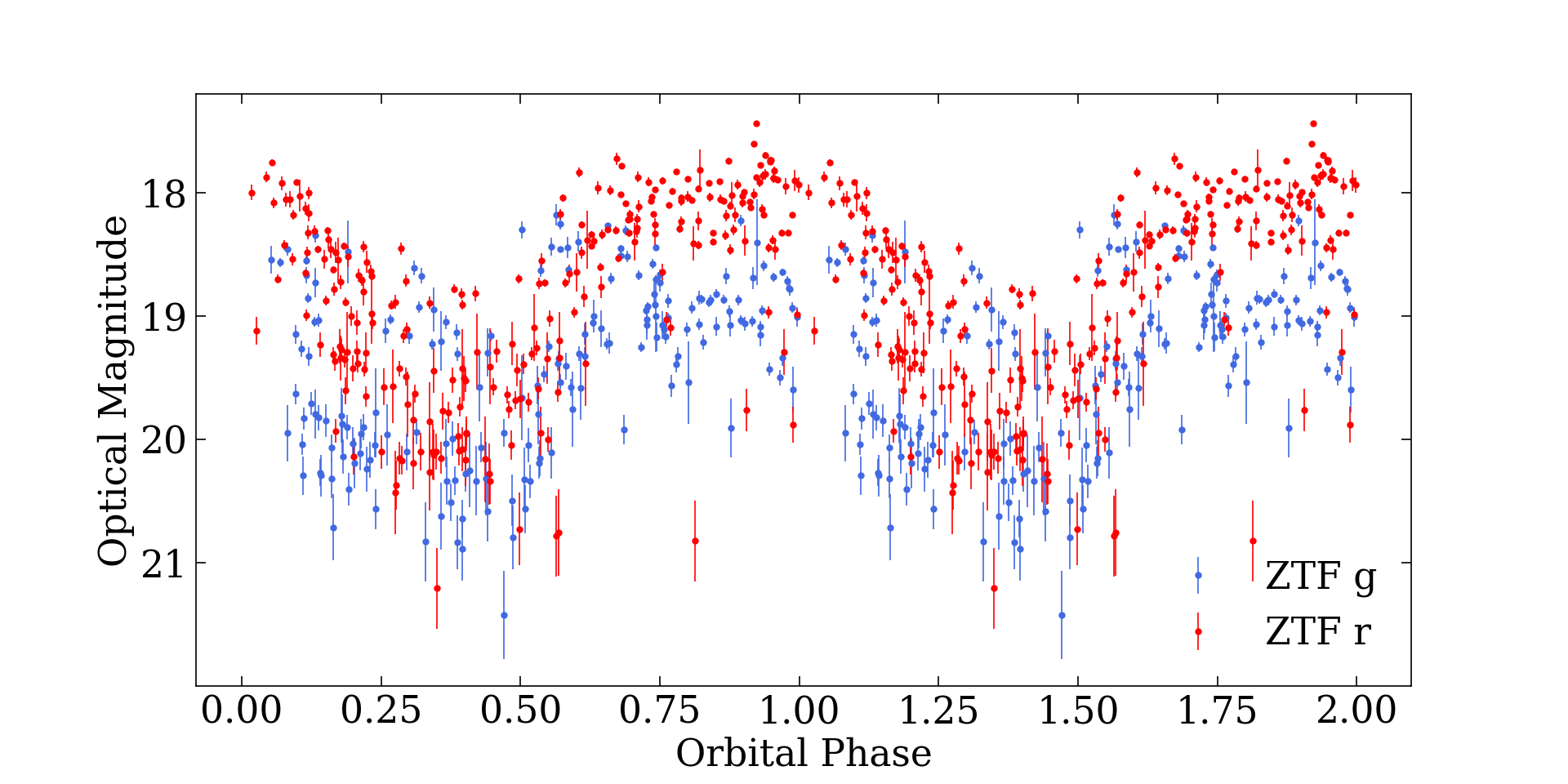}
    \caption{Folded lightcurve of ZTFJ0850+0443 (top, $P_\textrm{orb} = 1.72$ hr) over ZTF forced photometry. Large amplitude variations (1--2 mag) are characterstic of cyclotron beaming in polars.}
    \label{fig:ztf18_ZTF}
\end{figure}

ZTFJ0850+0443 was found in both ZTF $r$ and $g$ band data to be a periodic source with high amplitude. The best-fit Lomb-Scargle period is 103.44 minutes (1.72 hr) in both bands. The folded lightcurve is shown in Figure \ref{fig:ztf18_ZTF}. 

We followed up ZTFJ0850+0443 with high-cadence photometry in $r$ and $g$ bands using the Caltech HIgh-speed Multi-color camERA \citep[CHIMERA;][]{chimera} and $u$ band using the Wafer-Scale Imager for Prime \citep[WASP;][]{wasp}. The WASP $u$-band data were only acquired over a single period, while the CHIMERA $r$-and $g$-band data were both acquired simultaneously over two orbital periods. The high-cadence data revealed the eclipse in ZTFJ0850+0443 as well as cyclotron beaming at two points during a single orbit as seen in Figure \ref{fig:ztf18_large}.

An identification spectrum of ZTFJ0850+0443 was acquired using the Dual Imaging Spectrograph on the 3.5-m Apache Point telescope on 03 January 2022. That spectrum revealed strong He II 4686 compared to the H$\beta$ Balmer line. This is strong evidence of a magnetic CV, although non-magnetic CVs can sometimes show this line behavior \citep{1992silber, 2020oliveiraII}. Additional spectra at three orbital phases were obtained on the Keck telescope using the Low-Resolution Imaging Spectrometer \citep[LRIS;][]{lris} on 01 February 2022. A full orbit of ZTFJ0850+0443 was acquired on 07 March 2022 using LRIS. A summary of all data acquired for ZTF0850+0443 is presented in Table \ref{tab:data}.

\begin{deluxetable*}{cccc}
 \tablehead{
 \colhead{Data Type} & \colhead{Date} & \colhead{Instrument} & \colhead{Finding}  }
 \tablecaption{Data Acquired for ZTFJ0850+0443}
 \startdata
Identification Spectrum & 08 Jan. 2022 & {\parbox{4cm}{\centering \vspace{5pt} Apache Point 3.5-m Telescope/DIS}} & {\parbox{4cm}{\vspace{5pt}\centering Strong He II indicative of magnetic nature}}\\
High-cadence $u$ band Photometry & 27 Jan. 2022 & {\parbox{4cm}{\vspace{5pt}\centering Hale Telescope/WASP }}& {\parbox{4cm}{\vspace{5pt}\centering Eclipse and Cyclotron Beaming Revealed }}\\
Multi-phase spectra & 01 Feb. 2022 &{\parbox{4cm}{\vspace{5pt}\centering Keck I/LRIS}} &{\parbox{4cm}{\vspace{5pt}\centering Emission/Absorption Line Reversals and Multi-Component Emission}}\\
High-cadence $r$ and $g$ band Photometry & 04 Feb. 2022 & {\parbox{4cm}{\vspace{5pt}\centering Hale Telescope/CHIMERA}} & {\parbox{4cm}{\vspace{5pt}\centering High-cadence photometry at simultaneous orbital phases}}\\
Multi-phase spectra & 07 Mar. 2022 &{\parbox{4cm}{\vspace{5pt}\centering Keck I/LRIS}} & {\parbox{4cm}{\vspace{5pt}\centering Full Orbit for Doppler Tomography and Radial Velocities\vspace{5pt}}}\\
\enddata
\end{deluxetable*}
\label{tab:data}

\label{sec:results}

\subsection{General Lightcurve Features}

\begin{figure*}
    \centering
    \includegraphics[scale=0.45]{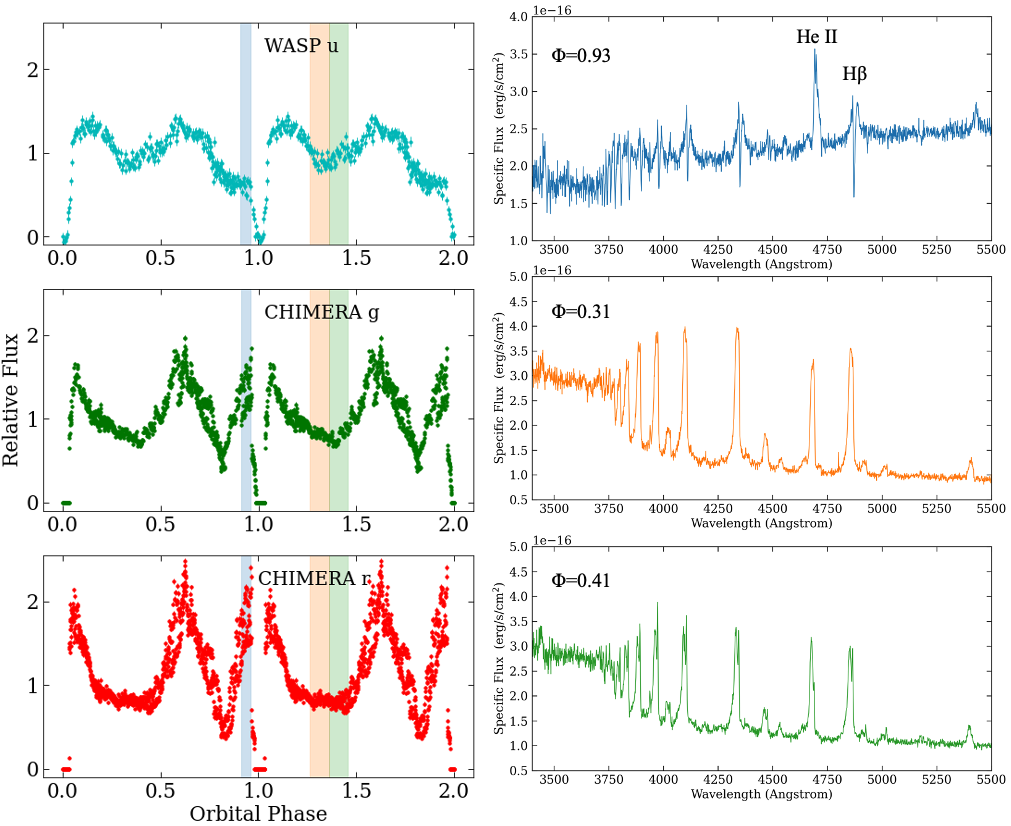}
    \caption{High-cadence photometry taken over two orbits (left) and multi-phase spectroscopy (right) of ZTFJ0850+0443. The highlights on the left panels correspond to the spectrum of the same color on the right. At phase $\phi = 0.93$, pre-eclipse line inversion is seen. At phase $\phi = 0.31$, the emission lines begin to split, revealing the irradiated face of the secondary star. At $\phi = 0.41$, the emission lines completely split and the irradiated .}
    \label{fig:ztf18_large}
\end{figure*}

In Figure \ref{fig:ztf18_large} we present the high-cadence photometry and spectroscopy of ZTF0850+0443 during notable orbital phases. The most prominent features of the high-cadence lightcurve are 1) the eclipse and 2) the two broad bumps per orbital phase in the $u$, $g$, and $r$ bands, one of which occurs around the eclipse and the other at phase $\phi \approx 0.6$.

\begin{figure}
    \centering
    \includegraphics[scale=0.3]{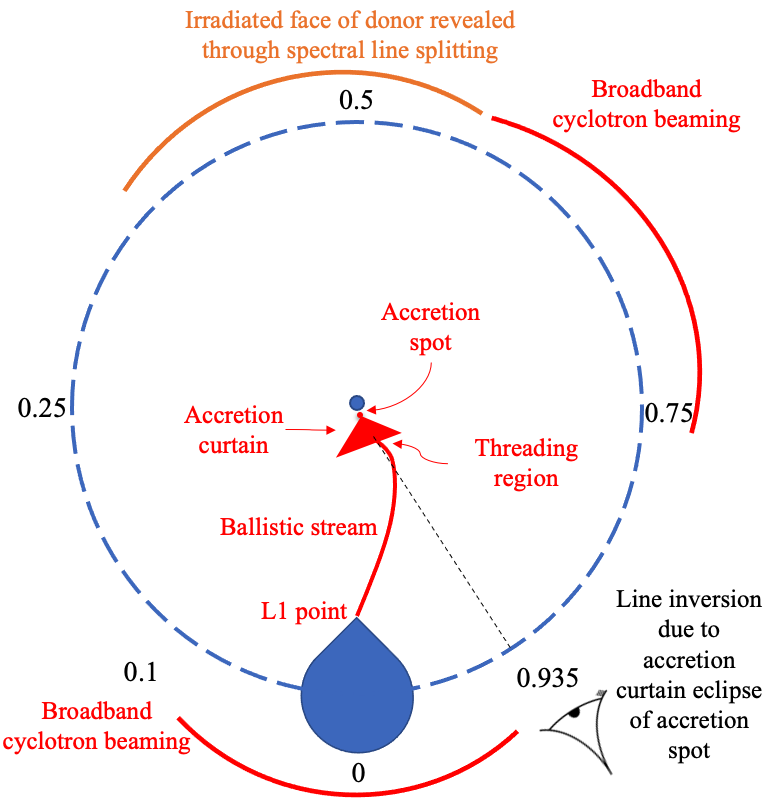}
    \caption{Cartoon of the orbit of ZTF0850+0443, based on Figure 12 of \cite{schmidt2005}. The observer can be imagined as rotating along the dashed circle as a function of orbital phase. }
    \label{fig:cartoon}
\end{figure}

As described in the Introduction, the magnetic field in polars channels accreted material directly from the threading region. The material is directed by the magnetic field out of the orbital plane through the ``accretion curtain" and onto the WD surface via one or two magnetic poles; see Figure 11 of \cite{littlefield2018}. Furthermore, the strong magnetic fields in polars lead to beamed cyclotron emission as non-relativistic spiral around magnetic field lines (synchrotron emission is the relativistic analog). Cyclotron emission can manifest itself in optical lightcurves as broad 1--2 mag bumps \citep[e.g.][]{1990cropper, hellierbook}. The data on ZTFJ0850+0443 points to it being an eclipsing polar. A visual aid to describe the system configuration is presented in Figure \ref{fig:cartoon}. 

In most polars ($\approx$ 95 percent), the WD spin is locked with the orbit. We searched to see if ZTFJ0850+0443 is an ``asynchronous" polar (i.e. the WD spin period is not locked to the orbital period). After Gaussian smoothing of the high-cadence lightcurves, no separate WD spin period is seen, suggesting that ZTFJ0850+0443 is a (typical) tidally locked polar. For the most part, orbital periods of polars range between 1.5 and 4 hours, placing the 1.72 hr period of ZTFJ0850+0443 well within the range of most polars \citep[e.g.][]{2018halpern, pala2020, 2020abril}.

In the following subsections, we walk through the orbit of ZTFJ0850+0443 and incorporate photometric and spectroscopic data to support our analysis.

\subsection{Pre-Eclipse}
\begin{figure}
    \centering
    \includegraphics[scale=0.32]{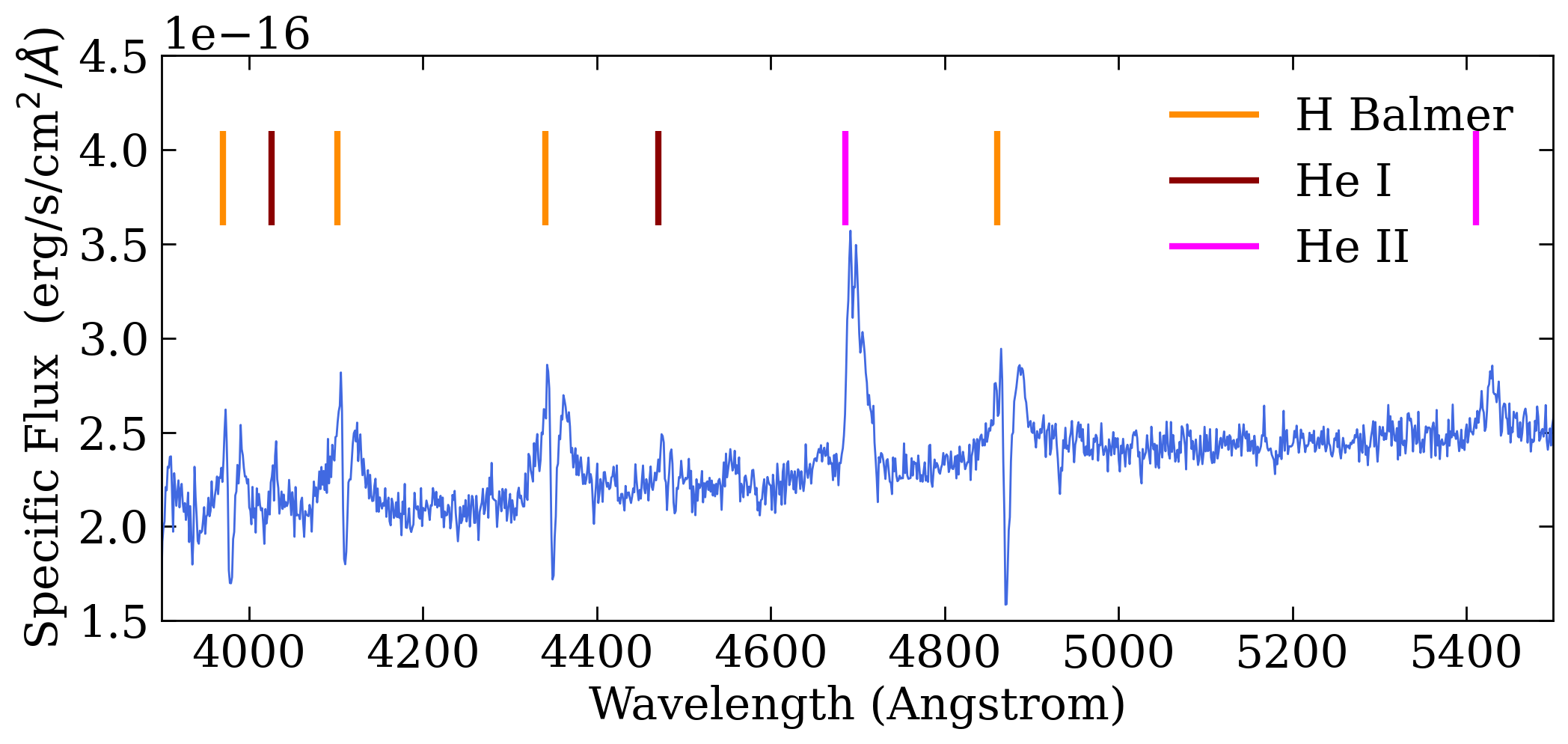}
    \caption{Zoomed-in spectrum of ZTFJ0850+0443 at $\phi =0.93$. He II is less broad than H Balmer lines and hardly absorbed. H Balmer lines are absorbed redward of line center.}
    \label{fig:ztf18_zoomed}
\end{figure}

CVs, both magnetic and non-magnetic, typically feature strong emission lines due to accretion. ZTFJ0850+0443 is one of six polars \citep{littlefield2018} where line inversions are seen in the pre-eclipse phase. Figures \ref{fig:ztf18_large} and \ref{fig:ztf18_zoomed} show that at phase $\phi = 0.93$, the H Balmer and He I lines are absorbed redward of line center. The He II 4686 line is only slightly absorbed. This also the case in the non-eclipsing polar MASTER OT J132104.04+560957.8 \citep{littlefield2018} as well as the eclipsing polar FL Cet (SDSS J015543.40+002807.2) \citep{schmidt2005}. \cite{littlefield2018} attribute this phenomenon to absorption within the accretion curtain (i.e. within the magnetosphere), not the threading region as was suspected before MASTER OT J132104.04+560957.8 was discovered. The main reason for this is that threading region is confined to the orbital plane, while the accretion curtain comes out of the plane along with the white dwarf magnetic field lines. Since ZTFJ0850+0443 is an eclipsing system, the presence of line inversions supports the idea that this phenomenon is more common at higher inclinations ($i\rightarrow 90^\circ$).

Another notable feature seen at pre-eclipse is that the $u$-band flux drops while the $r$- and $g$-band flux increase (Figure \ref{fig:ztf18_large}). This is due to the $u$ passband being centered blueward of the Balmer jump. In the pre-eclipse occultation of accretion spot emission by the accretion curtain, cooler material passes in front of the dominant source of radiation. Thus, the Balmer jump transitions from emission to absorption.  

\subsection{Eclipse: A Likely One-Pole System}
\label{sec:eclipse}

\begin{figure}
    \centering
    \includegraphics[scale=0.35]{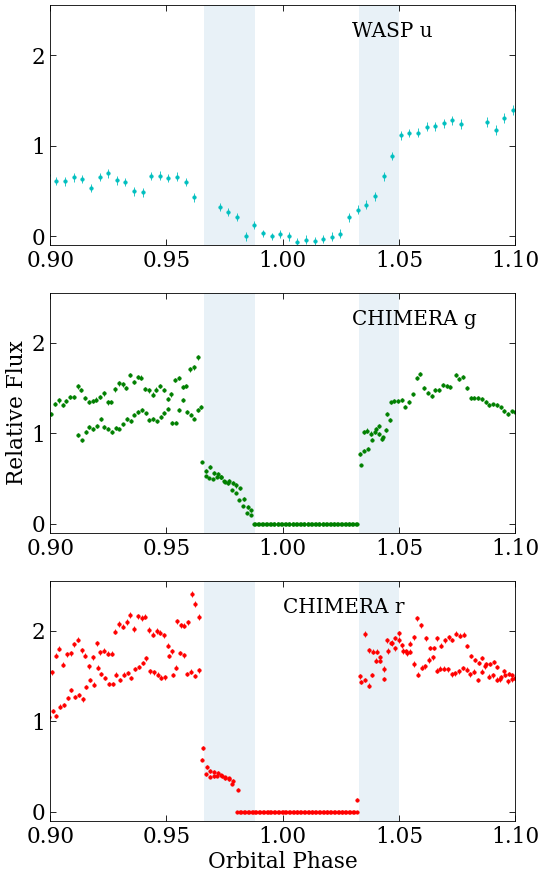}
    \caption{High-cadence lightcurve of ZTFJ0850+0443 around the eclipse. In the one-pole accretion model, the accretion spot on the white dwarf is eclipsed by the donor star at $\phi=0.966$. The accretion stream is gradually eclipsed until totally disappearing at $\phi=0.988$. The accretion spot exits eclipse at $\phi=1.033$ and the accretion stream gradually re-emerges until being fully exposed at $\phi= 1.05$.}
    \label{fig:eclipse}
\end{figure}

The lightcurve near eclipse can often be used to identify whether a polar has one or two accreting poles. We present the high-cadence lightcurve zoomed in around the eclipse in Figure \ref{fig:eclipse}. The data favors ZTFJ0850+0443 being a one-pole system, but we cannot discard the possibility of a second low-luminosity accreting pole. 

In the single-pole case, the pole is eclipsed at $\phi=0.966$ as indicated by the sharp decline in the $g$- and $r$-band data. The sampling of the $u$-band data is insufficient to resolve this. The first shaded region ($\phi=0.966 - 0.988$) in Figure \ref{fig:eclipse} corresponds to the gradual ingress of the ballistic stream. The pole exits eclipse at $\phi=1.033$ as indicated by the sharp rise in the $g$- and $r$-band data. The second shaded window ($\phi=1.033 - 1.05$) shows the gradual egress of the accretion curtain. We denote $\phi=1.05$ as the end of the accretion stream eclipse as that is the point where the total flux returns to its pre-eclipse value in $r$- and $g$-band data. The $u$-band flux is larger post-eclipse due to the absence of pre-eclipse self-absorption by the accretion curtain. Previous observations and numerical simulations have been able to reproduce remarkably similar lightcurves \citep{2019ecl_model, 2019ecl_model2}. 

In the case of two-pole accretion, two ``steps" in the ingress and egress of the magnetic poles are often seen. However, our photometric sampling cadence (10 seconds) is too low to detect that. For example, a $\sim 0.1$-second cadence was needed to see this in FL Cet \citep{2006salticam} and a 1-second cadence for eRASStJ192932.9--560346 \citep{schwope2021}. It could also be that the second pole is so faint at optical wavelengths that it would not appear in optical photometry at all. After all, no spectroscopic evidence aside from weak evidence in the Doppler tomograms (see Section \ref{sec:doptom}) of a second pole is seen. Additional data (e.g. optical polarimetry, X-ray lightcurves, or higher cadence optical photometry) are needed to definitively classify this as a one- or two-pole system, although the current data more strongly support the one-pole model. 

On the whole, the eclipse is shallowest in the $u$ band and deepest in the $r$ band. This has been attributed to the accreted material being channeled in the lowest energy configuration, which concentrates cooler material towards the center of the accretion stream. Hotter material is therefore more sparse and can still be seen while the cooler, concentrated material is eclipsed \citep[see Figure 5 of][]{1999huaquarii} for the modeling of HU Aquarii). We note that in the case of HU Aquarii, modeling of this feature alone in optical lightcurves was insufficient to determine whether the system is accreting at one or two poles.

\subsection{Donor Star Revealed}
\begin{figure}
    \centering
    \includegraphics[scale=0.31]{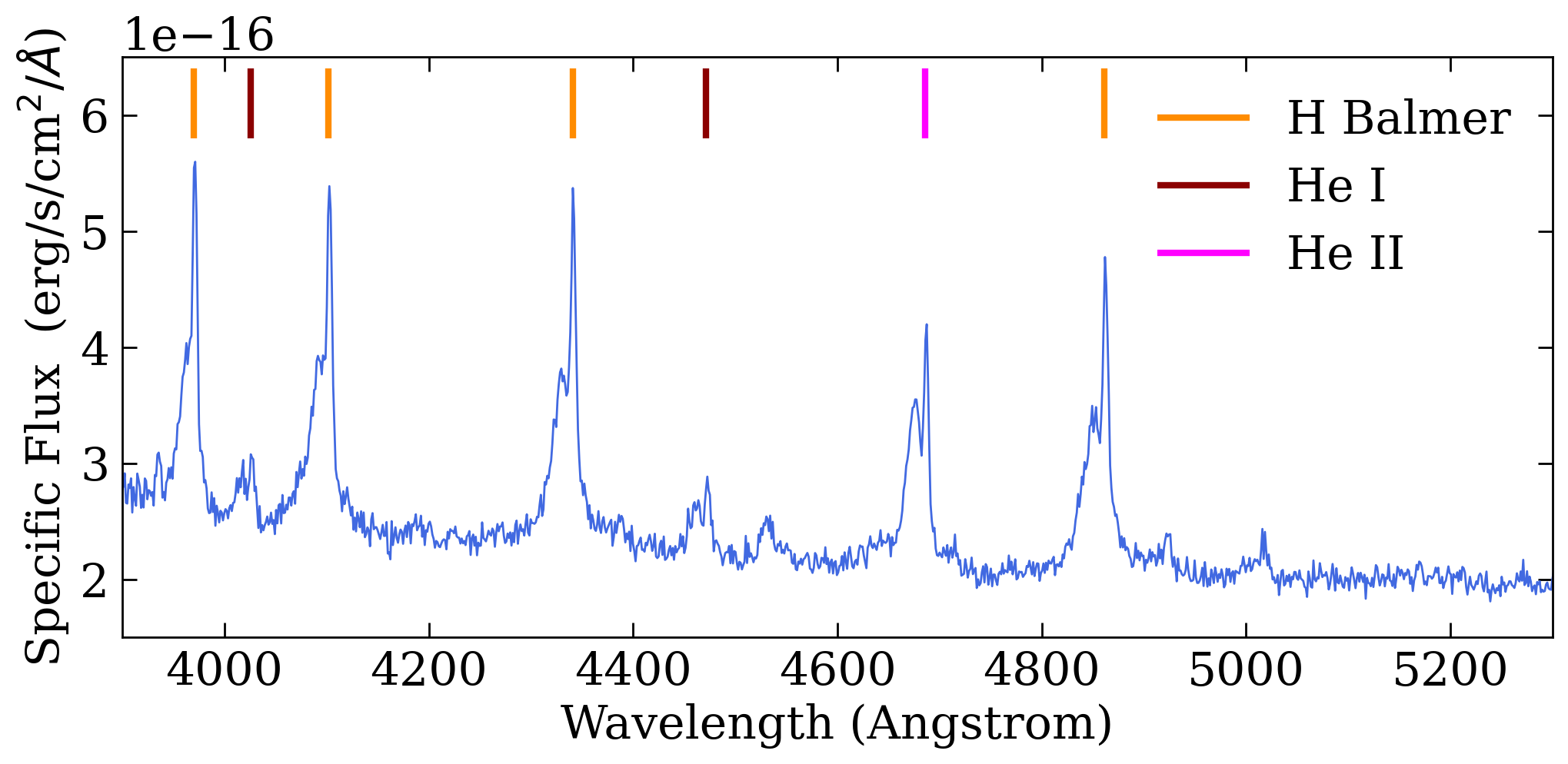}\\
    \includegraphics[scale=0.31]{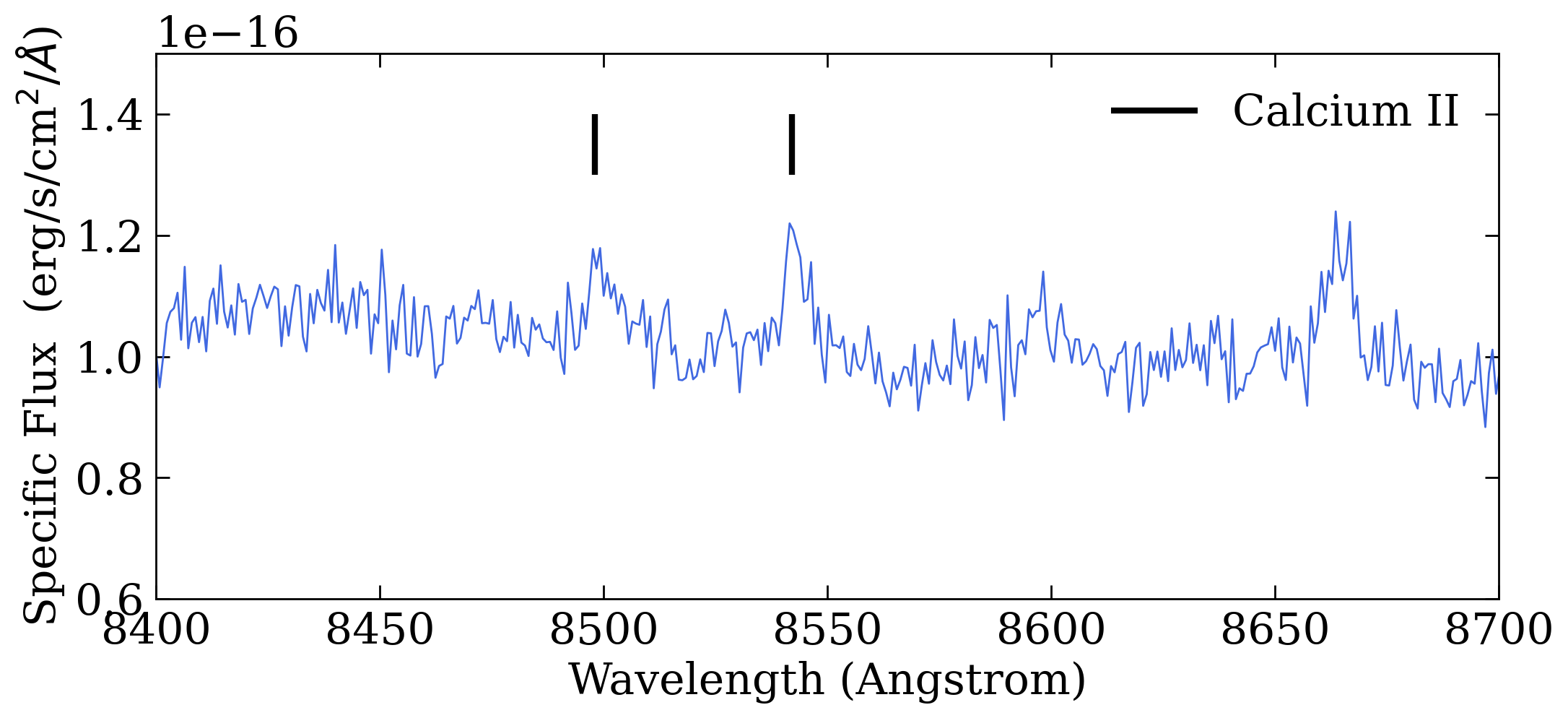}
    \caption{The narrow component of H Balmer and He I/II emission lines roughly trace the tip of the Roche lobe of the donor star (top). Ca II lines trace the irradiated face at the center-of-light (bottom). Spectra shown are taken near $\phi=0.5$.}
    \label{fig:secondary}
\end{figure}

At phase $\phi \approx 0.3$, the H Balmer, He I, and He II emission lines begin to split before being distinguishably separated at $\phi \approx 0.4$ (Figure \ref{fig:ztf18_large}). The broad, blueshifted component traces the accretion onto the white dwarf. The narrow component, which at this phase is slightly redshifted, traces the irradiated face of the donor star. At phase $\phi \approx 0.5$, the narrow component in the H Balmer, He I, and He II emission lines is at line center and stronger than the blueshifted accretion spot emission lines (Figure \ref{fig:secondary}).

At this point in the orbital phase, we are seeing directly into the irradiated face of the donor, confirming the origin of the narrow emission component. This phenomenon has been known since early studies of polars \citep[e.g.][]{1990cropper} and used to constrain binary parameters of the eclipsing polar BS Tri \citep{2022bstri} when the traditionally used Na I 8183, 8195 doublet could not be spotted. The Na I 8183, 8195 doublet is not seen in ZTFJ0850+0443 due to the strong accretion continuum that dominates out to red optical wavelengths. At the same orbital phases that H Balmer, He I, and He II emission lines are split into broad and narrow components (center around $\phi = 0.5$), Ca II 8498 and 8542 from the irradiated donor are seen in emission above the accretion continuum (Figure \ref{fig:secondary}). 

Na I, Ca II, H Balmer, He I, and He II lines from the donor star do not all arise from the same location. The H Balmer, He I, and He II lines likely arise from the tip of the Roche lobe and possibly into the accretion stream \citep[e.g.][]{2011schwope}. The Ca II lines roughly trace the center-of-light of the irradiated donor, somewhere between the center-of-mass and the tip of the Roche Lobe. The Na I lines tend to arise from deeper within the donor and more reliably trace the center-of-mass \citep[e.g.][]{2011schwope}. Therefore, in order to correctly use the Ca II lines to trace the radial velocity of the donor, we must apply a correction which we explain in detail in Section \ref{sec:binary_param}.

\subsection{Doppler Tomography}
\label{sec:doptom}
\begin{figure}
    \centering
    \includegraphics[scale=0.5]{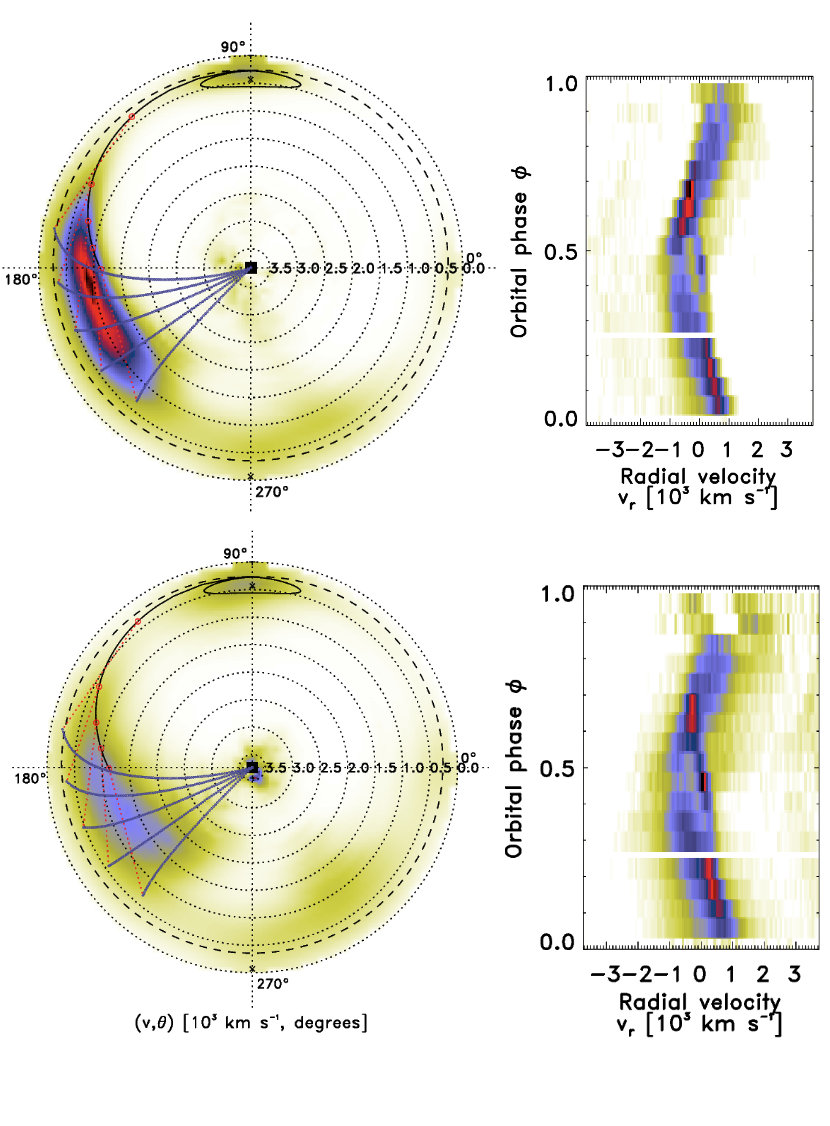}
    \caption{``Inverse" Doppler tomograms and radial velocity curves for ZTF0850+0443 using He II 4686 (top) and H$\beta$ (bottom). Redder color indicates stronger line strength. Model magnetic field threading is shown for clarity using generic parameters. The dominant component is that of the accretion curtain. The irradiated face of the secondary can also be seen towards the top.}
    \label{fig:doptom}
\end{figure}

Spectral lines in CVs vary as a function of orbital phase, often containing information about all system parameters which can be blended together. In CVs with an accretion disk, the disk, the accretion hot spot (where the accretion stream hits the disk), and irradiated face of the donor star can all contribute to the observed emission. Doppler tomography converts phase-resolved spectroscopy into a plot of observed radial velocity and line strength as a function of orbital phase. Doppler tomograms disentangle the contribution of the various CV components (e.g. accretion disk, accretion hot spot, donor star) to a given spectral line; see \cite{doptomography} for a review of the method of Doppler tomography. We present Doppler tomogams and radial velocity curves of He II 4686 and H$\beta$ in Figure \ref{fig:doptom}.

We use the \texttt{doptomog}\footnote{\url{https://www.saao.ac.za/~ejk/doptomog/main.html}} code developed by \cite{2015kotze}. We show the ``inverse" Doppler tomograms, which better illustrate the high-velocity components in magnetic CVs. Higher velocities are located closer in to the center of the diagram, while lower velocities are farther from center (traditional Doppler tomography flips this around).

The feature with an amplitude of 1000 km/s is the dominant component. The free-fall velocity at the surface of a white dwarf, assuming typical parameters ($M_\textrm{WD}= 0.6 M_\odot, R_\textrm{WD}= R_\textrm{Earth}$) is $\approx$ 4000 km/s. By observing emission at 1000 km/s, we can infer that this emission is due to material within the accretion curtain as it approaches the WD surface. We do not know enough about the magnetic field configuration near the surface to know exactly where in the accretion curtain we are probing. A weak, diffuse component can also be seen towards the bottom right, which could be indicative of a second magnetic pole. This is the only possible evidence for a second pole that we have in our current data. 

The irradiated face of the secondary is clearly shown roughly between 300--400 km/s in both Doppler tomograms as well near $\phi=0.5$ in the radial velocity curves. The gap near $\phi=0.3$ is due to a small gap in our data acquisition. The radial velocity curves clearly show the pre-eclipse line inversion, where He II 4686 is hardly split, but H$\beta$ is split into two components due to the intervening redshifted absorption.

\subsection{Binary Parameters}
\label{sec:binary_param}


In order to solve for the full mass and radius parameters of the system, we assume a circular orbit. Our constraining equations are the following: 1) the Roche lobe equation from \cite{eggleton83}, 2) the binary star mass function, 3) the relationship between the mass ratio and the correction applied to the RVs derived from Ca II lines, 4) an $R(M)$ relation derived from modern CV evolutionary tracks, and 5) the eclipse of the system.

The Roche lobe equation \citep{eggleton83} is:
\begin{gather}
    \frac{R_L}{a} = \frac{0.49 q^{2/3}}{0.6 q^{2/3} + \ln\left(1 + q^{1/3}\right)} = f(q)
\end{gather}
where $a$ is the orbital separation of the system: $a^3 = G(M_1 + M_2)P_\textrm{orb}^2 / 4\pi^2$ and $q$ is the ratio of the donor star mass to the white dwarf mass: $q = M_2/M_1$. We adopt this mass ratio convention consistently throughout this study.

The binary star mass function is:
\begin{gather}
    \frac{(M_1\sin i)^3}{(M_1 + M_2)^2} = \frac{P_\textrm{orb}K_2^3}{2\pi G}
    \label{eq:binary_eqn}
\end{gather}

where $M_1$ is the mass of the accreting white dwarf and $M_2$ is the mass of the donor. $K_2$ is the radial velocity of the donor, which we must infer from the observed radial velocity of the Ca II lines, $K_2'$. As discussed earlier, these lines originate from the center-of-light of the irradiated (day) side of the donor star, which is not a good approximation for the center-of-mass of the donor star. We obtain $K'_2=360 \pm 15 \textrm{ km/s}$ from a least-squares fit to the observed line profiles (Figure \ref{fig:rv}). The statistical error of 15 km/s arises from instrumental precision and low signal-to-noise in measuring the line position at certain phases. We assume the entire donor star is co-rotating around the center of mass of the system at the same orbital period. To determine the relationship between $K_2$ and $K_2'$, we assume the center-of-light of the donor star is located $\varepsilon R_L$ from the center-of-mass of the donor star. The factor $\varepsilon$ can be thought as being the coordinate on the axis between the donor center-of-mass and the tip of the Roche lobe. Therefore, the semi-major axis of the center-of-light is $a_2' = a_2 - \varepsilon R_L$ and the observed radial velocity is $K_2' = 2\pi (a_2 - \varepsilon R_L)/P_\textrm{orb}$. Since $K_2 =  2\pi a_2/P_\textrm{orb}$, we can solve for the correction factor: $K_2'/K_2 = 1 -\varepsilon R_L/a_2$. Since $a = a_2(1 + q)$, we write:
\begin{equation}
    \frac{K_2'}{K_2} = 1 - \varepsilon f(q)\times(1+q)
\end{equation}
\label{eq:correction}
and therefore the RV correction is a function of the mass ratio $q = M_1/M2$ and $\varepsilon$, the coordinate on the axis between the donor center-of-mass and the tip of the Roche lobe. A value of $\varepsilon = 0$ corresponds to the donor center-of-light being located at the donor center-of-mass and a value of $\varepsilon = 1$ corresponds to the donor center-of-light being located at the tip of the Roche lobe.

In order to determine reasonable values of $\varepsilon$, we constructed a simple binary star model using \texttt{PHOEBE} \citep{2005phoebe1, 2016phoebe2} consisting of a 15,000 K primary star \citep[average of WDs in CVs at this orbital period, e.g.][]{pala2022} irradiating a 3000 -- 4000 K donor star at a typical CV orbital separation. This model with a single source of radiation serves as the lower limit of possible radiation for the donor. In polars, radiation from the accretion stream, accretion curtain, and accretion pole can also irradiate the donor. In our simple model, the center-of-light (location of mean intensity) is at $\varepsilon=0.45$, which serves as a lower limit.

We then turn to empirical findings to place an upper limit on $\varepsilon$. We consider the analysis of He II, Ca II, and Na I lines of the donor star in HU Aquarii \citep{2011schwope}. Doppler tomography of those lines showed that the Ca II lines precisely probe the center-of-light, located approximately at values of $\varepsilon=0.5-0.85$. 

\begin{figure}
    \centering
    \includegraphics[scale=0.29]{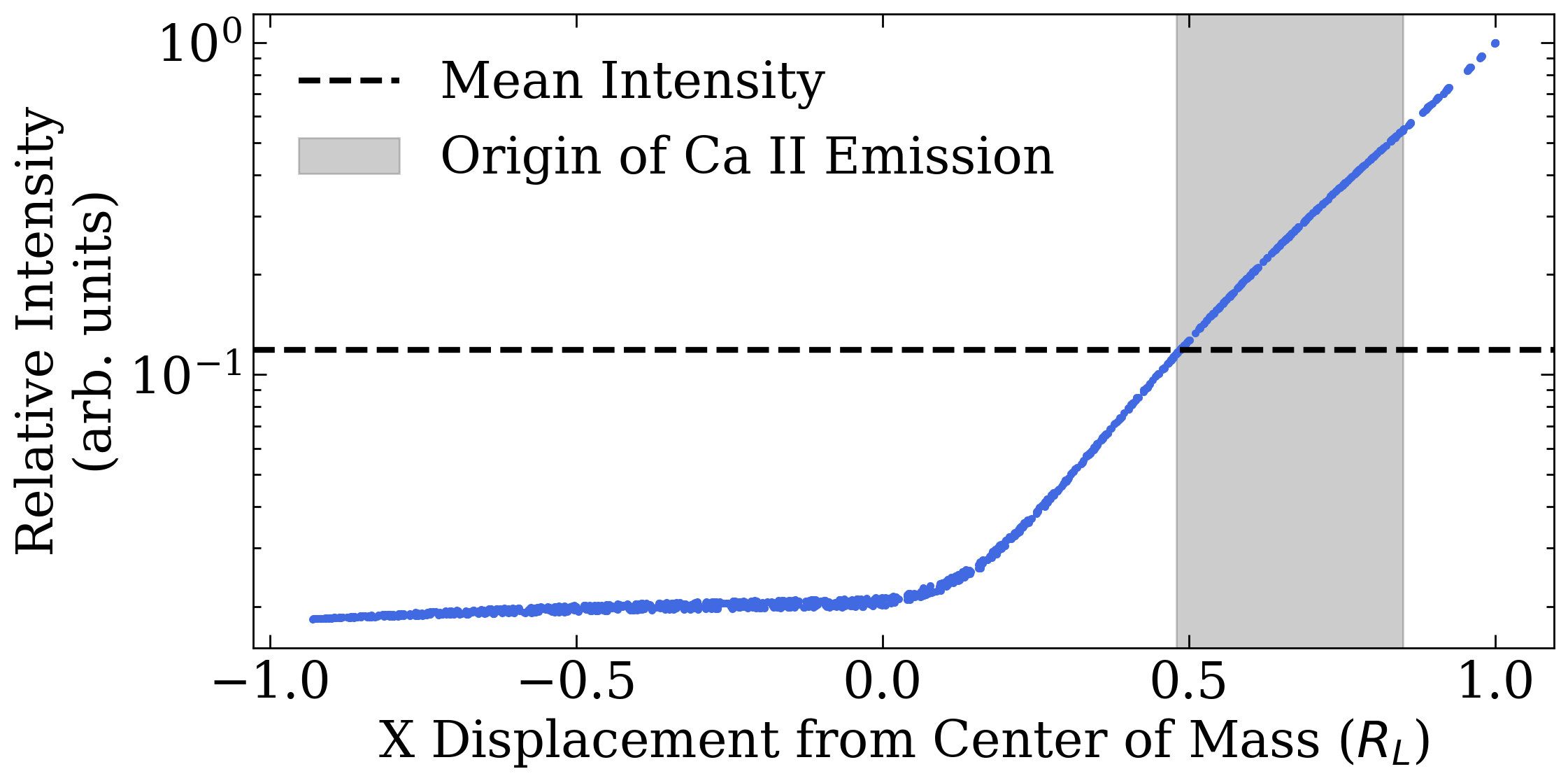}
    \caption{Output of a \texttt{PHOEBE} model of a late-type donor being irradiated by a 15,000K white dwarf. The mean intensity (center-of-light) is located at $\varepsilon=0.45$, while an empirical analysis of the donor in a similar polar places the origin of Ca II emission lines to be as far away from the center-of-mass as $\varepsilon=0.85$.}
    \label{fig:phoebe}
\end{figure}

Therefore, we adopt the range $\varepsilon=0.45-0.85$, combining the simple analytical model and empirical results. We overplot this range on our \texttt{PHOEBE} model in Figure \ref{fig:phoebe}. A full treatment of radiation from the white dwarf, accretion spot, accretion curtain, accretion stream, and efficiency of irradiation of the donor are beyond the scope of this study.

\begin{figure}
    \centering
    \includegraphics[scale=0.25]{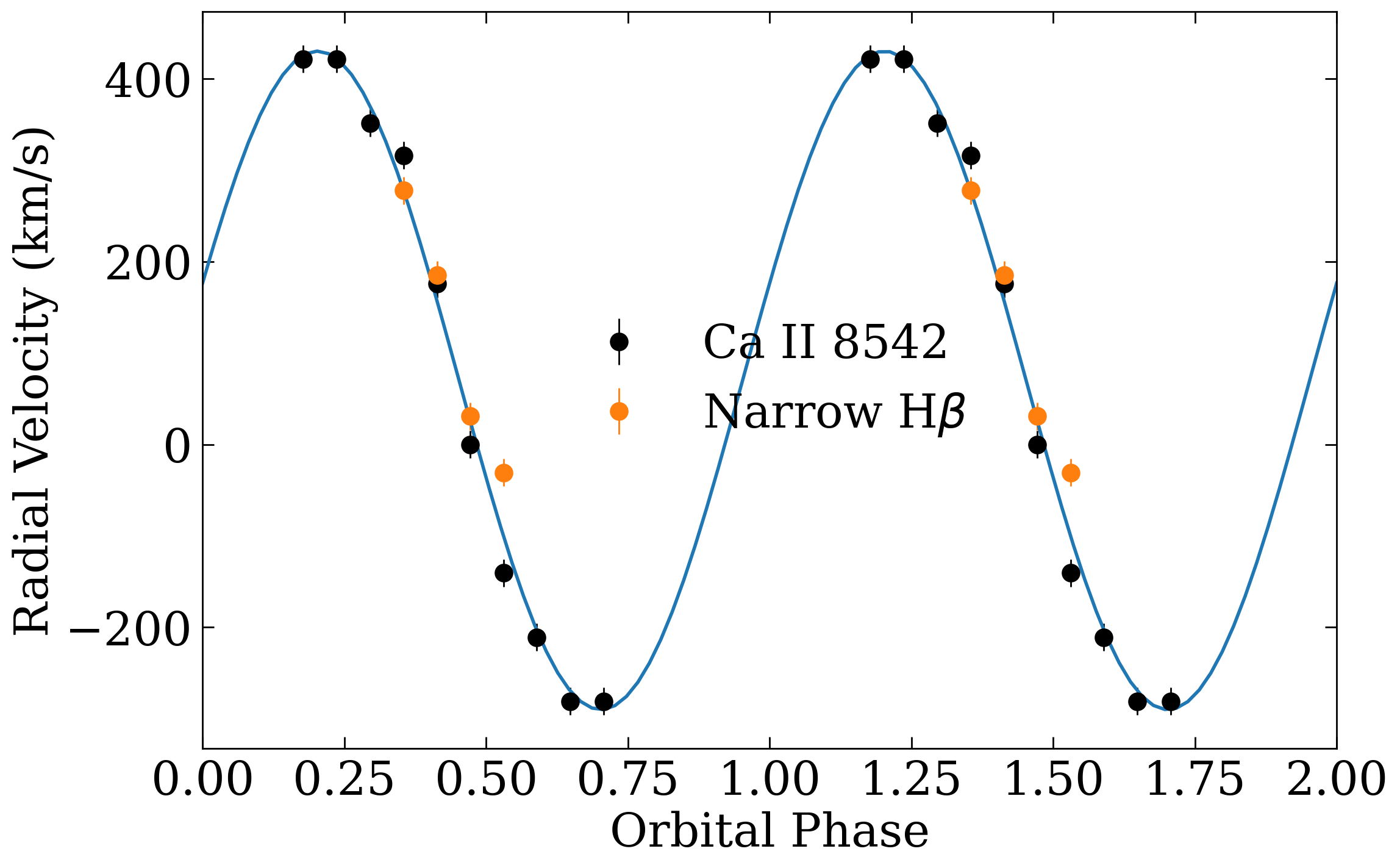}
    \caption{Radial velocity (RV) measurements of the irradiated face of the donor star as a function of orbital phases,  $K_2'$. RV measurements are obtained at orbital phases where 1) Ca II is strong enough to be seen above the continuum or 2) the narrow component of H Balmer lines is not blended with the broad accretion component.}
    \label{fig:rv}
\end{figure}

In order to obtain an $R(M)$ relation, we use the CV donor star evolutionary tracks from the \cite{2011knigge} set of models. They account for the inflated radius of the donor star due to the rotation it undergoes while in the binary system.

The final component is the eclipsing nature of the system. Given the typical system parameters of polars, we know the inclination of the system must be $i \gtrsim 78^\circ$ for the system to be eclipsing. \cite{1976chanan} showed that if the eclipse timing is well-constrained, then the mass ratio $q$ is a function of the inclination $i$ in Roche lobe filling systems. We find the eclipse duration to be 415 $\pm$ 5 sec, measured at the sharp dropoff and sharp rise as described in Section \ref{sec:eclipse}.

The \cite{1976chanan} geometry, however, assumes a point source located at the center of the white dwarf and not on the surface, as is that case in polars. This however, leads to a negligible systematic that is contained within the statistic error bars on the measurement of all system parameters.

We then solve for all equations simultaneously via a vectorized least-squares approach implemented with \texttt{scipy} to find the final best-fit values. This is equivalent to minimizing the error in Equation \ref{eq:binary_eqn}, Equation \ref{eq:correction}, $R_L = R(M_2)$, and the eclipse timing equation from \cite{1976chanan} simultaneously. We present the final values for the binary system in Table \ref{tab:parameters}.

\begin{deluxetable}{cc}
 \tablehead{
 \colhead{System Parameter} & \colhead{Estimated Value} }
 \tablecaption{System Parameters for ZTFJ0850+0443}
 \startdata
$M_1\;(M_\odot)$ & $0.81 \pm 0.08$\\
$M_2 \;(M_\odot)$ & $0.119 \pm 0.002$\\
$R_2 \;(R_\odot)$ & $0.163 \pm 0.002$\\
$K_2 \;(\textrm{km/s})$ & $434 \pm 15$\\
$i$ (degrees) & $83.3^\circ \pm 1.2^\circ$\\
$P_\textrm{orb}$ (hr) & 1.724\\
$\Delta t_\textrm{ecl}$ (sec) & 
$415 \pm 5
$\\
$\dot{M}$ ($M_\odot$/yr) & $\sim 10^{-11}$\\
$\varepsilon = (a_2-a_2')/R_L
$ & $0.65 \pm 0.20$
\enddata
\end{deluxetable}
\label{tab:parameters}


The largest uncertainty in our white dwarf mass estimate stems from the uncertainty in radial velocity (which scales to the third power in Kepler's binary system equation). While this is limited by our estimates of the location of the donor center-of-light (parameterized by $\varepsilon$), we find that we can still reasonably constrain the mass of the white dwarf.

\subsection{Magnetic Field Strength}

There are three common ways to determine the magnetic field strength of a polar as outlined by \cite{1990cropper}: 1) Cyclotron humps in the optical spectrum, 2) Zeeman splitting of emission lines in the optical spectrum, 3) Optical polarization measurements. In the case of ZTF0850+0443, we do not detect the first two and have not acquired optical polarization data to test the third criterion. 

Why don't we see any cyclotron harmonics? The wavelengths of cyclotron harmonics \citep[e.g.][]{2015ferrario, 2019lowfieldpolar} are given by the following formula:
\begin{gather}
    \lambda_n = \frac{10710}{n}\left(\frac{100 \textrm{ MG}}{B}\right)\sin\theta\; \textup{\AA}
\end{gather}
where $\lambda_n$ is in Angstrom and $\theta$ is the viewing angle of the cyclotron beaming. In cases where cyclotron humps are seen in the optical, it is usually the $n=3$ or $n=4$ harmonics that are seen \citep{1990cropper, 2000oldmagrev}. Since lower-order harmonics are less prominent, the absence of seeing cyclotron harmonics at optical wavelengths often implies low magnetic field strength. \cite{1993ferrario_stlmi} and \citep{2000oldmagrev} showed that in polars with field strengths of $B \lesssim 20$ MG, cyclotron harmonics ($n=2,3,4$) can only be seen in near-infrared spectra and not at all in optical spectra.  

\begin{figure}
    \centering
    \includegraphics[scale=0.31]{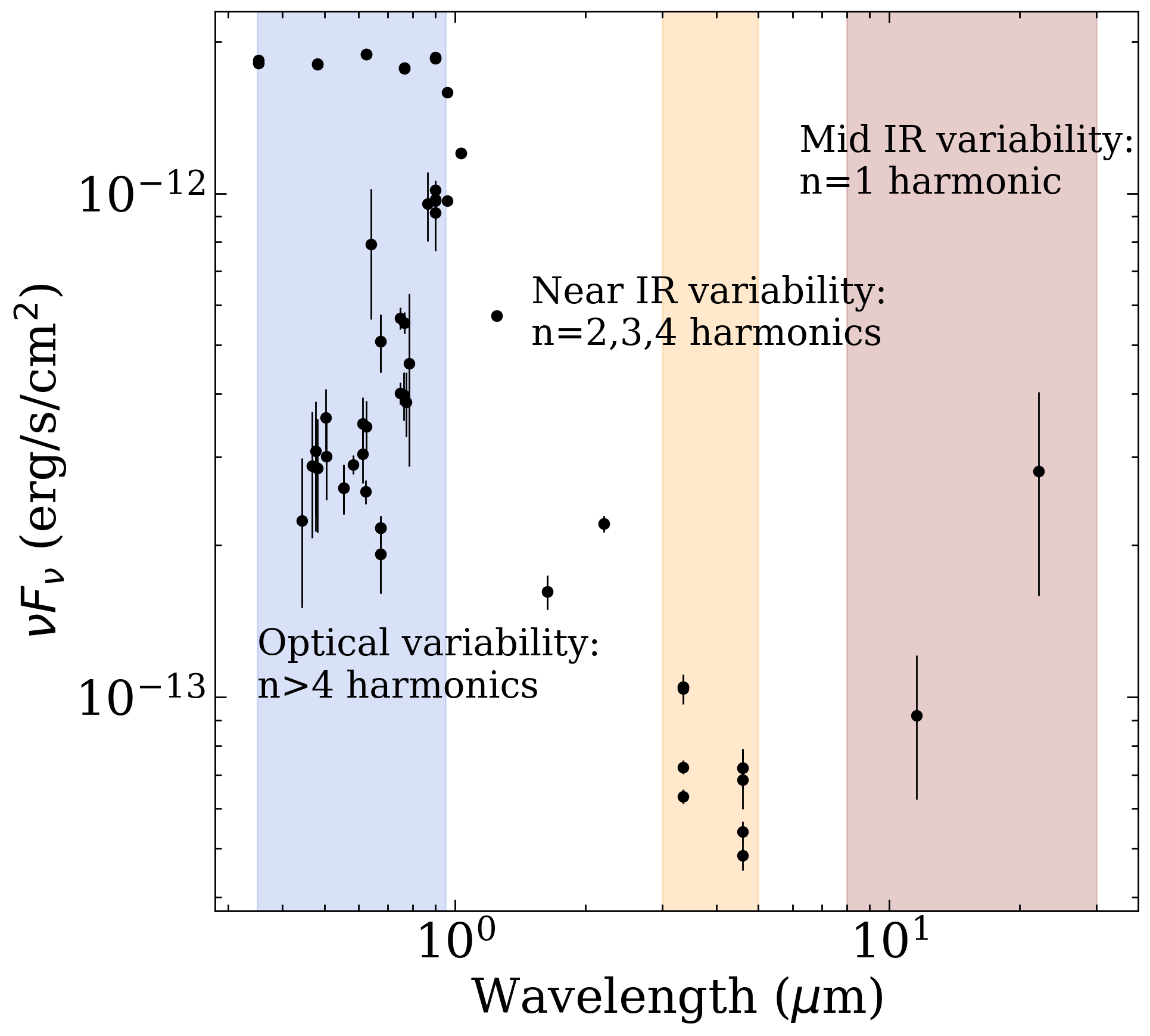}
    \caption{Spectral energy distribution of ZTFJ0850+0443 reproduced from CDS. Variability is seen at optical wavelengths. The WISE W3 and W4 points are at 11.6 and 22.1 $\mu$m, respectively. The clear mid-infrared excess of ZTFJ0850+0443 suggests it is a low-field ($B \lesssim 10$ MG) polar. }
    \label{fig:sed}
\end{figure}

When infrared spectra are not available, infrared photometry has been shown to be insightful for magnetic field characterization of low-field polars \citep{2015wisepolars, 2000oldmagrev}. \cite{2019lowfieldpolar} discovered CRTS J035010.7 +323230, a polar that did not show any of the typical magnetic field diagnostics outlined by \cite{1990cropper}.  CRTS J035010.7 +323230 stood out by showing a strong infrared excess with WISE W3 and W4 magnitudes higher than W2 and W1 points. The authors argued this was indicative of a low-field ($\lesssim 10$ MG) polar. We see exactly the same behavior in ZTF0850+0443, with a clear infrared excess in the WISE W3 and W4 bands. We reproduce a spectral energy distribution in Figure \ref{fig:sed} compiled from all data in CDS which includes photometry from SDSS, PANSTARRS, Gaia, WISE, GALEX, and UKIDSS, among others surveys.

We propose that ZTF0850+0443 must be viewed at a high enough angle ($\sin \theta \sim 1$) from the magnetic pole so that cyclotron beaming can be observed in its optical lightcurve. What we see in the optical is likely a blend of many low-order cyclotron harmonics in the lightcurve bumps at phase $\phi = 0.6$ and $\phi = 1$. The fundamental ($n=1$) harmonic is then in the mid-infared ($\sim 10-20 \mu$m) with strong $n=2,3,4$ harmonics in the near-infrared. We suggest that ZTF0850+0443 is a low-field ($B \lesssim 10$ MG) polar, potentially adding to a small pool of polars with such low magnetic field strengths.

\subsection{Accretion Rate}
We can estimate the accretion rate of ZTF0850+0443 assuming conversion from potential energy to X-ray luminosity. Assuming a $\sim$10 percent efficiency of this energy conversion \citep[e.g.][]{2007rosswog}:
\begin{gather}
    L_X \sim 0.1\times \frac{1}{2}\frac{G M_\textrm{WD}\dot{M}}{R_\textrm{WD}}
\end{gather}
\label{eq:cyclotron}
Assuming the standard white dwarf mass-radius relation \citep[e.g.][]{hellierbook}, for an X-ray flux of $2\times 10^{31}$ erg/s and a mass of $M_\textrm{WD} \approx 0.81 M_\odot$, we obtain $\dot{M} \sim 10^{-11} M_\odot$/yr.

\section{ZTFJ0926+0105}
\label{sec:0926}

\subsection{Data}
\begin{figure}
    \centering
    \includegraphics[scale=0.13]{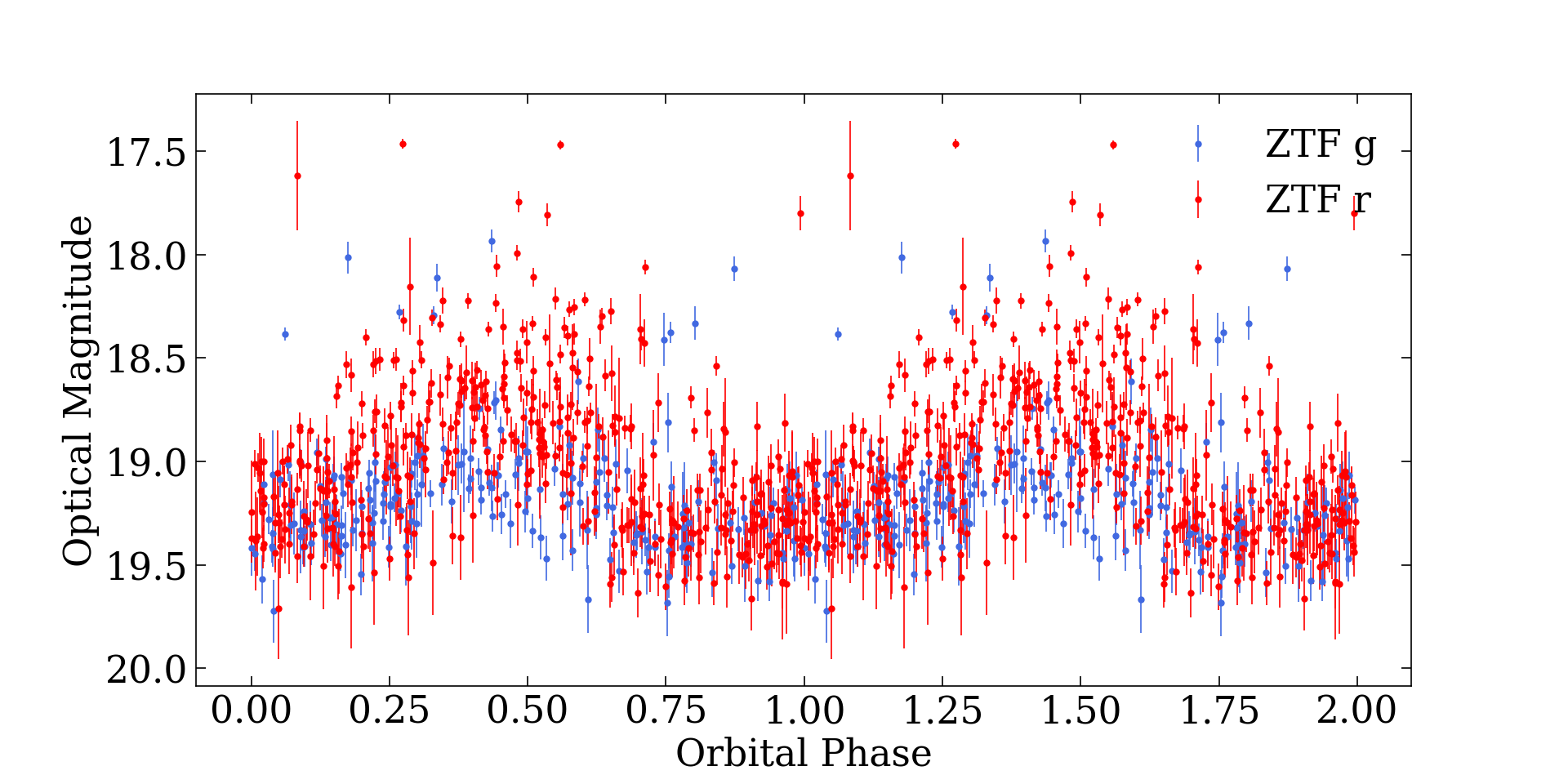}
    \caption{Folded lightcurve of ZTF0926+0105 ($P_\textrm{orb} = 1.48$ hr) over ZTF forced photometry. }
    \label{fig:ztf17_ZTF}
\end{figure}

ZTFJ0926+0105 was found in  ZTF data to be a periodic source with high amplitude in both $r$ and $g$ bands (Figure \ref{fig:ztf17_ZTF}). We followed up ZTFJ0926+0105 with high-cadence photometry using CHIMERA. Data were acquired over 3.5 hours, but all orbital phases could not be completely covered due to transient clouds. The best-fit period of the $r$-band data is 88.53 min, while the best-fit period of the $g$-band data is 89.10 min. The average of the two is 88.81 min (1.48 hr).

An identification spectrum was acquired using the Double Spectrograph \citep[DBSP;][]{dbsp} on the Hale telescope on 08 January 2022. This spectrum showed strong H Balmer, He I, and He II lines in emission. He II 4686 fulfilled the criteria of \cite{1992silber} for being a magnetic CV candidate, prompting us to acquire follow-up spectroscopy.  Phase-resolved spectra were taken on the Keck telescope using the Echelle Spectrograph and Imager \citep[ESI;][]{esi} on 01 February 2022. Eight 10-minute spectra were taken consecutively with $\sim$11 minutes between the starting point of each spectrum. Table \ref{tab:data_0926} summarizes all data taken and the contribution of each dataset.

\begin{deluxetable*}{cccc}
 \tablehead{
 \colhead{Data Type} & \colhead{Date} & \colhead{Instrument} & \colhead{Finding}  }
 \tablecaption{Data Acquired for ZTFJ0926+0105}
 \startdata
Identification Spectrum & 07 Jan. 2022 & {\parbox{4cm}{\centering \vspace{5pt} Hale Telescope/DBSP}} & {\parbox{4cm}{\vspace{5pt}\centering Strong He II indicative of magnetic nature}}\\
High-cadence $r$ and $g$ band Photometry & 05 Feb. 2022 & {\parbox{4cm}{\vspace{5pt}\centering Hale Telescope/CHIMERA}} & {\parbox{4cm}{\vspace{5pt}\centering High-cadence photometry at simultaneous orbital phases}}\\
Multi-phase spectra & 06 Feb. 2022 &{\parbox{4cm}{\vspace{5pt}\centering Keck II/ESI}} & {\parbox{4cm}{\vspace{5pt}\centering Full Orbit for Doppler Tomography and Radial Velocities\vspace{5pt}}}\\
\enddata
\end{deluxetable*}
\label{tab:data_0926}

\subsection{Lightcurve and Spectral Analysis with Doppler Tomography}

\begin{figure}
    \centering
    \includegraphics[scale=0.21]{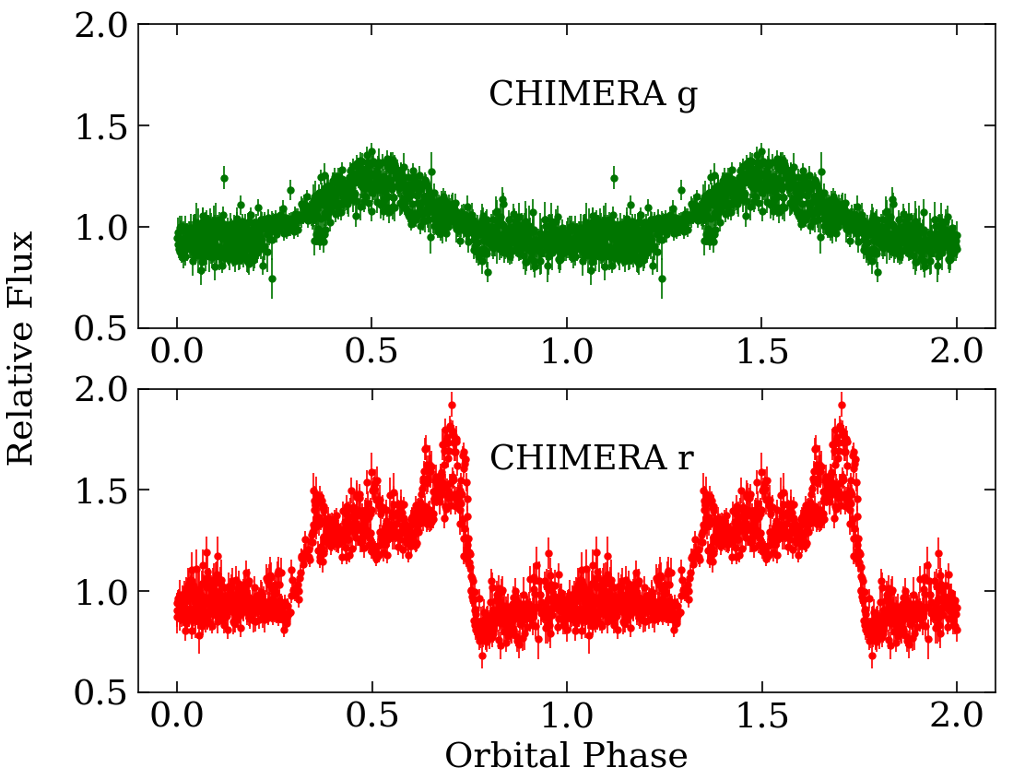}\\
    \includegraphics[scale=0.26]{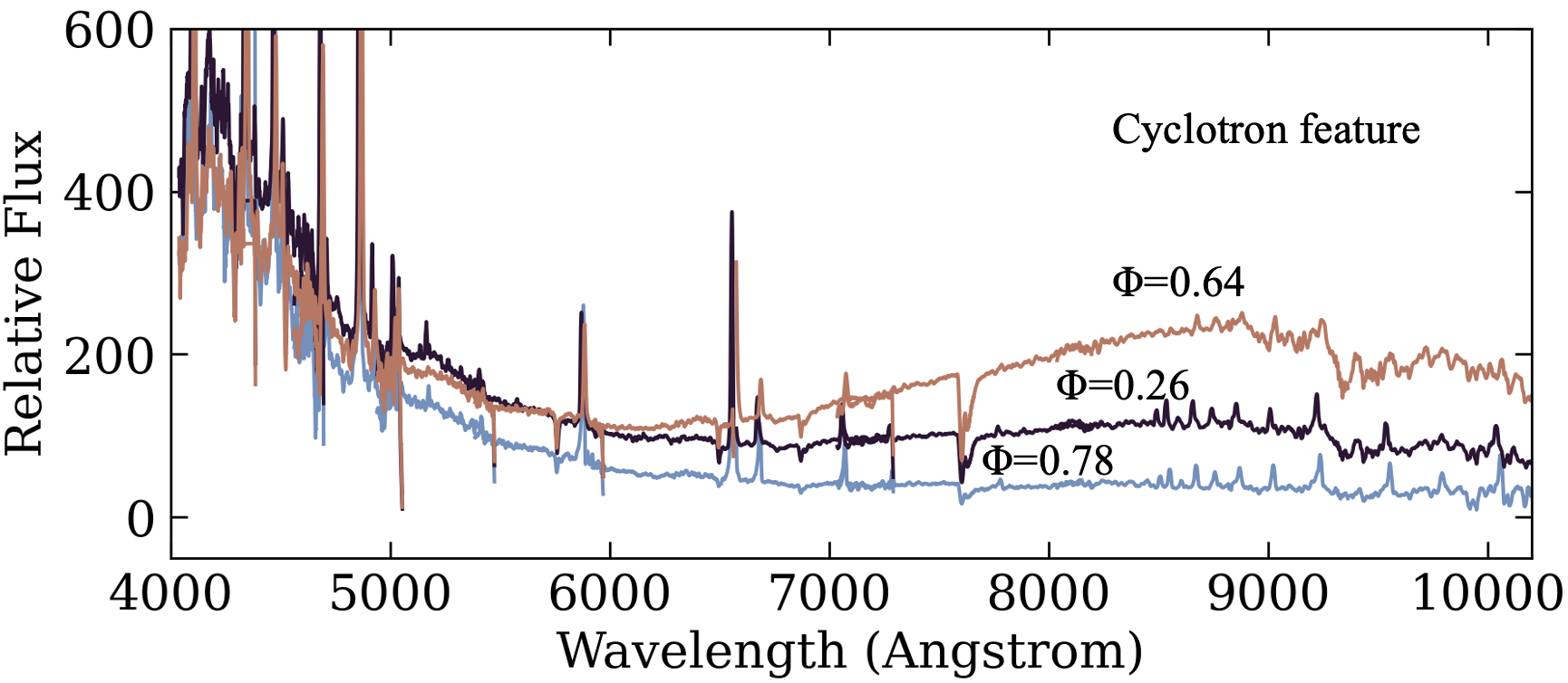}
    \caption{Cyclotron beaming is seen once per orbital phase in the high-cadence CHIMERA photometry (top) as well as phase-resolved spectroscopy around 8800 $\textup{\AA}$ (bottom).}
    \label{fig:0926_chimera}
\end{figure}

We see cyclotron beaming once per orbital phase in the $g$ and $r$ high-cadence lightcurves of ZTFJ0926+0105 (Figure \ref{fig:0926_chimera}). The large factor by which the flux increases, combined with the orbital period of 1.48 hr, is characteristic of few objects other than polars. No eclipse or any other notable features are seen. We also see a prominent cyclotron bump in the phase-resolved spectra of ZTFJ0926+0105 (Figure \ref{fig:0926_chimera}). The phase where the bump is most prominent coincides with the maximum in the photometry, confirming cyclotron beaming as the source of the high-amplitude variation. We only see one cyclotron harmonic around 8800 Angstrom, and discuss the implications for determining the magnetic field strength in Section \ref{sec:bfield_0926}.

Aside from the prominent cyclotron bump, the phase-resolved spectroscopy of ZTFJ0926+0105 is much like that of ZTFJ0850+0443, revealing emission from both the accretion curtain and the face of the irradiated donor star. We produce Doppler tomograms for ZTFJ0926+0105 in order to disentangle the two components. Both the He II 4686 and H$\beta$ tomograms reveal large, approximately $750$ km/s amplitude radial velocities. We attribute such large radial velocities to the accretion curtain, as was the case with $ 1000$ km/s radial velocities in ZTFJ0850+0443. Since ZTFJ0926+0105 is viewed at a more face-on inclination, the observed radial velocity should be lower.

\begin{figure}
    \centering
    \includegraphics[scale=0.45]{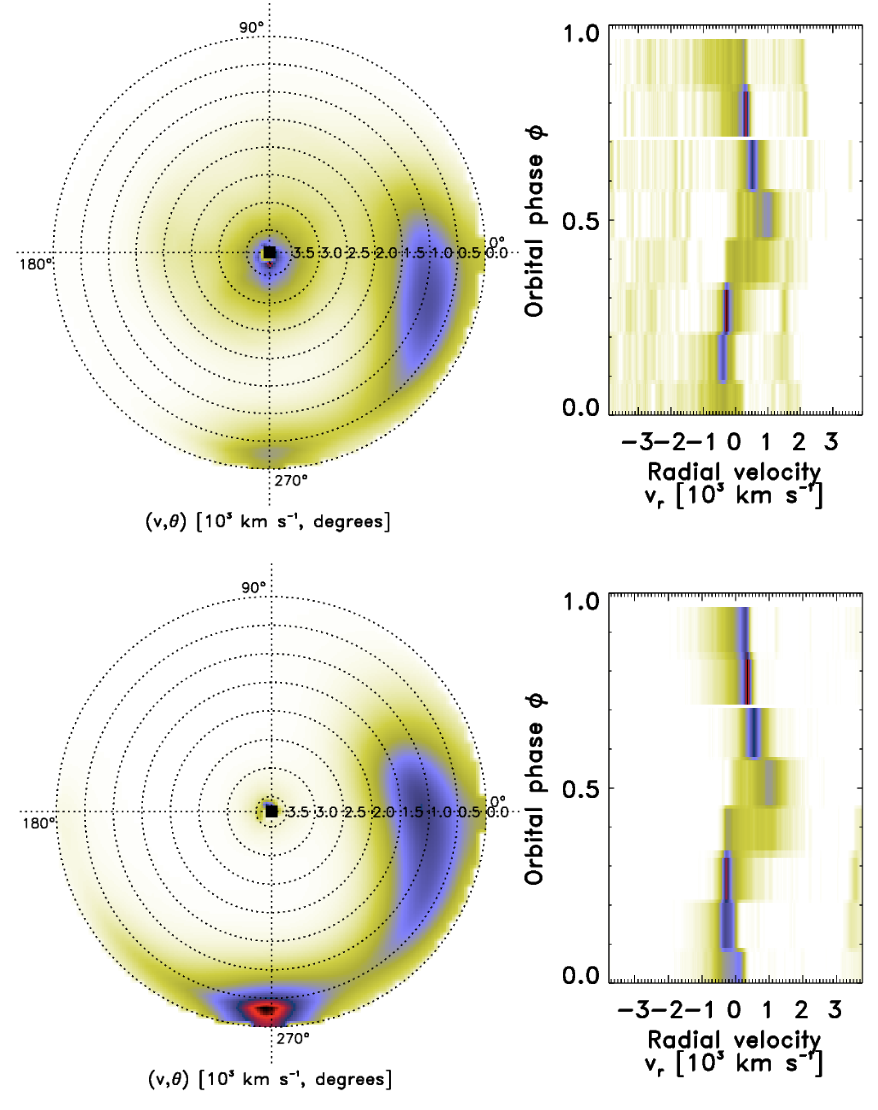}
    \caption{``Inverse" Doppler tomograms and radial velocity curves for ZTF0926+0105 using He II 4686 (top) and H$\beta$ (bottom). Redder color indicates stronger line strength. The dominant component is that of the accretion curtain. The irradiated face of the secondary can also be seen at the bottom, appearing stronger in H$\beta$ compared to He II 4686. }
    \label{fig:doptom_0926}
\end{figure}

Both Doppler tomograms of ZTFJ0926+0105 (Figure \ref{fig:doptom_0926}) reveal the irradiated face of the donor star at approxmiately 300 km/s, particularly the H$\beta$ tomogram. This coincides with the presence of Ca II lines (see Figure \ref{fig:0926_chimera}) which also stem from the irradiated face of the donor star like in ZTFJ0850+0443. While we could estimate radial velocity measurements of the donor star in ZTFJ0926+0105, the lack of an eclipse prevents us from obtaining a precise estimate of the white dwarf mass. 

\subsection{Magnetic Field Strength and Accretion Rate}
\label{sec:bfield_0926}

We can at best put a lower limit on the magnetic field strength of ZTF0926+0105. We are primarily limited by the resolution and flux calibration of our data at blue optical wavelengths. In the phase-resolved spectroscopy, we see a single prominent cyclotron feature at phase $\phi = 0.68$. This harmonic is well-fit by a Gaussian centered at 8800 Angstrom. 

There are no other cyclotron harmonics seen at phase $\phi = 0.68$, where the maximum cyclotron continuum should occur. After performing a power-law continuum fit, subtracting the single cyclotron feature and possible template spectra of the donor, no other cyclotron harmonics bumps emerge. The bump-like features at 6400 Angstrom and 5100 Angstrom are in the spectra at all phases, suggesting they are not cyclotron features which should be beamed only once per orbital phase. 

Using Equation \ref{eq:cyclotron}, we must assume both a viewing angle and cyclotron harmonic number to find the magnetic field strength. Such a strong cyclotron feature like the one we see at 8800 Angstrom is unlikely due to a harmonic higher than $n=4$ \citep[e.g.][]{1990cropper}. The viewing angle must also be large ($\th \gtrsim 60^\circ$) for the beaming to be as high amplitude as we see in the spectroscopy and photometry. We therefore place a lower limit of $B \gtrsim 26$ MG on the magnetic field strength of ZTF0926+0105. Near-infrared spectroscopy could be used to search for the lower harmonics and place stronger constraints on the field strength.

We can estimate the accretion rate of ZTF0926+0105 assuming conversion from potential energy to X-ray luminosity as we did with ZTF0850+0443. Adopting a $\sim$10 percent efficiency of this energy conversion and the standard white dwarf mass-radius relation \citep[e.g.][]{hellierbook}, for an X-ray flux of $4\times 10^{30}$ erg/s and a mass of $M_\textrm{WD} \approx 0.8$, we obtain $\dot{M} \sim 10^{-12} M_\odot$/yr.

\section{Discussion}
\label{sec:discussion}

\subsection{Placement on the HR Diagram}
\begin{figure}
    \centering
    \includegraphics[scale=0.4]{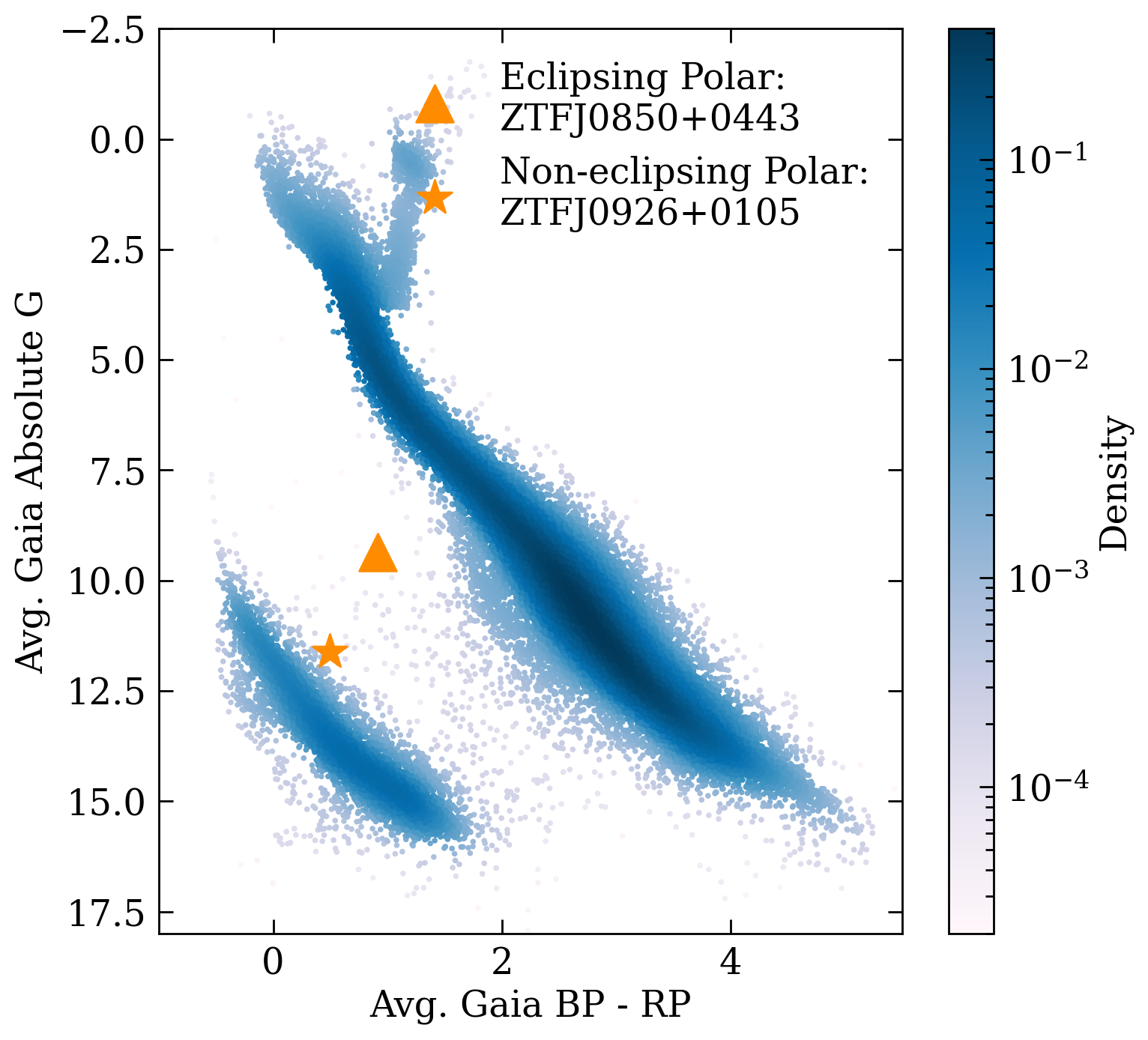}
    \caption{Gaia EDR3 HR diagram composed of sources within 100 pc with an exceptional parallax measurement ($\pi / \sigma/\pi > 10$). The three CVs in this work are significant outliers which reside between the main sequence and WD tracks.}
    \label{fig:gaia_hr}
\end{figure}

We overplot ZTFJ0850+0443 and ZTFJ0926+0105 on a Gaia Hertzsprung-Russell (HR) diagram (Figure \ref{fig:gaia_hr}). We note that ZTFJ0850+0443 has a parallax divided by parallax error ($\pi/\sigma_\pi$) value of 2.8. This is lower than the value adopted by some authors to claim a precise distance measurement ($\pi/\sigma_\pi > 3$ or even $\pi/\sigma_\pi > 10$, depending on the study). We plot all sources in Gaia EDR3 within 100 pc of the Sun with parallax $\pi/\sigma_\pi > 10$. On average, the locations of ZTFJ0850+0443 and ZTFJ0926+0105 are consistent with the mean position of polars as found by \cite{2020abril}.

\subsection{Current CV Population and SRG}
All CVs are X-ray emitters, to some extent \citep{2017mukai}. While magnetic CVs typically have higher X-ray luminosities than their non-magnetic counterparts, the eFEDS dataset is deep enough to reveal new X-ray detections of non-magnetic CVs. Table \ref{tab:cv_list} presents a list of all SDSS CVs in the eFEDS footprint alongside the objects reported in this paper. This list of CVs is by no means exhaustive, and there are CVs and CV candidates in the Open Catalysmic Variable Catalog \citep{opencvcatalog} that are not in the SDSS list. However, a complete analysis of all CVs in the eFEDS field is beyond the scope of this study.

Even from this modest sample, the efficiency of discovery of magnetic CVs is greatly increased by incorporating X-ray data from eFEDS/SRG. Neither of the two polars were listed as SDSS CVs, and ZTF0850+0443 was even previously misclassified as a quasar based on its GALEX color information \citep{2012galex_misclassification}. 

An all-sky eROSITA/SRG release will be vital for constructing a volume-limited survey similar to that of \cite{pala2020}. Volume-limited samples are one way of eliminating observational bias from constructing a full picture of CVs and their evolution, as was shown by \cite{pala2020}. Introducing X-ray information eliminates another observational bias and provides another property that can be used to study CVs. ZTFJ0850+0443 and ZTFJ0926+0105 were also detected in the ``hard" sample of the eFEDS catalog (2.3--5 keV) and fulfill at least one of the criteria to be classified as X-ray variables in the eFEDS data release \citep{salvato2021}.

\begin{deluxetable*}{c|c|c|c|c|c|c|c}
 \tablehead{
 \colhead{ZTF ID} & \colhead{RA (J2000)} & \colhead{DEC (J2000)}  & \colhead{Distance (pc)} & \colhead{$L_X$ (erg/s)} & 
 \colhead{$P_\textrm{orb}$ (hr)}  & \colhead{He II 4686?} & \colhead{Dwarf Novae?}
 }
 \tablecaption{List of SDSS CVs with $g <$ 21.5 mag in the eFEDS/ZTF Footprint}
 \startdata
ZTFJ0850+0443 & 132.6549698 & 4.732512195  & 1000 & $2 \times 10^{31}$ & 1.72 & Strong & No\\
ZTFJ0926+0105 & 141.5595715	 & 1.099278643 & 370 & $4 \times 10^{30}$ & 1.48 & Strong & No\\
SDSSJ0839+0223 & 129.947638 & 2.392059696	  & 140 & $2 \times 10^{28}$ & 2.58 & No & No\\
SDSSJ0847+0145 & 131.8974924 & 1.759408443 & - & - & - & Weak & No \\
SDSSJ0840+0005 & 130.1725288 & 0.0889426823  & - & - & - & No & No\\
SDSSJ0932+0109 & 143.1591464 & 1.150692234  & - & - & - & Weak & No\\
SDSSJ0843--0148 & 130.7645492 & -1.816258365   & 2200 & $7 \times 10^{31}$ & 4.15 & No & No\\
SDSSJ0856+0254 & 133.9618872 & -1.90794599  & 140 & $8 \times 10^{28}$ & - & No & No\\
SDSSJ0851+0308 & 132.78078791 & 3.1429164171  & 580 & $4 \times 10^{30}$ & - & Weak & Yes\\
SDSSJ0914+0137 & 138.5446499 & 1.62581610  & 1000 & $7 \times 10^{30}$ & - & No & Yes\\
SDSSJ0856+0254 & 134.0355156& 2.902766968 & 210 & $6 \times 10^{28}$ & - & No & No\\
\enddata

\end{deluxetable*}
\label{tab:cv_list}

\cite{schwope2021} identified an eclipsing polar through an eROSITA/SRG crossmatch with Gaia using a proprietary eRASS dataset. We showed here that the public eFEDS dataset revealed a similarly interesting object when crossmatched with ZTF.

Our findings show that the eROSITA/SRG X-ray survey is vital in supplementing ZTF for the discovery of new CVs. Making use of Gaia, we can obtain precise luminosities of the objects we find. The upcoming Gaia DR3 will be useful in this context.

\section{Conclusion}
We have discovered two polars: ZTFJ0850+0443 (eclipsing, $P_\textrm{orb} =1.72$ hr) and ZTFJ0926+0105 (non-eclipsing, $P_\textrm{orb} =1.48$ hr), through a crossmatch of the eFEDS dataset and ZTF archival photometry. We suggest that ZTFJ0850+0443 is likely a low-field polar with magnetic field $B_\textrm{WD} \lesssim 10$ MG. The accreting white dwarf in ZTFJ0850+0443 has a mass of $M_\textrm{WD} = 0.81 \pm 0.08 M_\odot$ and accretion rate of $\dot{M} \sim 10^{-11} M_\odot$/yr, both typical values of polars. ZTFJ0926+0105 has a more typical magnetic field strength of polars, $B_\textrm{WD} \gtrsim 26$ MG, and accretion rate of $\dot{M} \sim 10^{-12} M_\odot$/yr. Because it is not eclipsing, we cannot place robust measurements on the white dwarf mass.

This study is part of a larger follow-up analysis of the eFEDS/ZTF footprint. Studies such as this are useful in overcoming observational biases in previous optical-only searches for CVs, and will directly lead to accurate volume-limited studies of CVs such as that by \cite{pala2020}. This in turn will test our knowledge of the origin of magnetic fields in WDs, compact object accretion, and binary star evolution.

\section{Acknowledgements}
ACR thanks the ZTF Variable Star Group for useful comments and discussions. ACR also thanks Axel Schwope, Paul Groot, Frank Verbunt, and Jim Fuller for insightful conversations that led to an improved final manuscript. 

This work is based on observations obtained with the Samuel Oschin Telescope 48-inch and the 60-inch Telescope at the Palomar Observatory as part of the Zwicky Transient Facility project. Major funding has been provided by the U.S National Science Foundation under Grant No. AST-1440341 and by the ZTF partner institutions: the California Institute of Technology, the Oskar Klein Centre, the Weizmann Institute of Science, the University of Maryland, the University of Washington, Deutsches Elektronen-Synchrotron, the University of Wisconsin-Milwaukee, and the TANGO Program of the University System of Taiwan.

The ZTF forced-photometry service was funded under the Heising-Simons Foundation grant \#12540303 (PI: Graham).

Some observations were made with the Apache Point 3.5m telescope, which is owned and operated be the Astrophysical Research Corporation.

This work has made use of data from the European Space
Agency (ESA) mission Gaia (\url{https://www.cosmos.esa.int/gaia}), processed by the Gaia Data Processing and Analysis Consortium (DPAC, \url{https://www.cosmos.esa.int/web/gaia/ dpac/consortium}). Funding for the DPAC has been provided by national institutions, in particular, the institutions participating in the Gaia Multilateral Agreement.

This work is based on data from eROSITA, the soft X-ray instrument aboard SRG, a joint Russian-German science mission supported by the Russian Space Agency (Roskosmos), in the interests of the Russian Academy of Sciences represented by its Space Research Institute (IKI), and the Deutsches Zentrum für Luft- und Raumfahrt (DLR). The SRG spacecraft was built by Lavochkin Association (NPOL) and its subcontractors, and is operated by NPOL with support from the Max Planck Institute for Extraterrestrial Physics (MPE). The development and construction of the eROSITA X-ray instrument was led by MPE, with contributions from the Dr. Karl Remeis Observatory Bamberg \& ECAP (FAU Erlangen-Nuernberg), the University of Hamburg Observatory, the Leibniz Institute for Astrophysics Potsdam (AIP), and the Institute for Astronomy and Astrophysics of the University of Tübingen, with the support of DLR and the Max Planck Society. The Argelander Institute for Astronomy of the University of Bonn and the Ludwig Maximilians Universität Munich also participated in the science preparation for eROSITA.

\bibliography{main}{}

\begin{thebibliography}{}
\expandafter\ifx\csname natexlab\endcsname\relax\def\natexlab#1{#1}\fi
\providecommand{\url}[1]{\href{#1}{#1}}
\providecommand{\dodoi}[1]{doi:~\href{http://doi.org/#1}{\nolinkurl{#1}}}
\providecommand{\doeprint}[1]{\href{http://ascl.net/#1}{\nolinkurl{http://ascl.net/#1}}}
\providecommand{\doarXiv}[1]{\href{https://arxiv.org/abs/#1}{\nolinkurl{https://arxiv.org/abs/#1}}}

\bibitem[{{Abril} {et~al.}(2020){Abril}, {Schmidtobreick}, {Ederoclite}, \&
  {L{\'o}pez-Sanjuan}}]{2020abril}
{Abril}, J., {Schmidtobreick}, L., {Ederoclite}, A., \& {L{\'o}pez-Sanjuan}, C.
  2020, \mnras, 492, L40, \dodoi{10.1093/mnrasl/slz181}

\bibitem[{{Ag{\"u}eros} {et~al.}(2009){Ag{\"u}eros}, {Anderson}, {Covey},
  {Hawley}, {Margon}, {Newsom}, {Posselt}, {Silvestri}, {Szkody}, \&
  {Voges}}]{2009agueros}
{Ag{\"u}eros}, M.~A., {Anderson}, S.~F., {Covey}, K.~R., {et~al.} 2009, \apjs,
  181, 444, \dodoi{10.1088/0067-0049/181/2/444}

\bibitem[{{Bellm} {et~al.}(2019{\natexlab{a}}){Bellm}, {Kulkarni}, {Graham},
  {Dekany}, {Smith}, {Riddle}, {Masci}, {Helou}, {Prince}, {Adams},
  {Barbarino}, {Barlow}, {Bauer}, {Beck}, {Belicki}, {Biswas}, {Blagorodnova},
  {Bodewits}, {Bolin}, {Brinnel}, {Brooke}, {Bue}, {Bulla}, {Burruss}, {Cenko},
  {Chang}, {Connolly}, {Coughlin}, {Cromer}, {Cunningham}, {De}, {Delacroix},
  {Desai}, {Duev}, {Eadie}, {Farnham}, {Feeney}, {Feindt}, {Flynn},
  {Franckowiak}, {Frederick}, {Fremling}, {Gal-Yam}, {Gezari}, {Giomi},
  {Goldstein}, {Golkhou}, {Goobar}, {Groom}, {Hacopians}, {Hale}, {Henning},
  {Ho}, {Hover}, {Howell}, {Hung}, {Huppenkothen}, {Imel}, {Ip}, {Ivezi{\'c}},
  {Jackson}, {Jones}, {Juric}, {Kasliwal}, {Kaspi}, {Kaye}, {Kelley},
  {Kowalski}, {Kramer}, {Kupfer}, {Landry}, {Laher}, {Lee}, {Lin}, {Lin},
  {Lunnan}, {Giomi}, {Mahabal}, {Mao}, {Miller}, {Monkewitz}, {Murphy},
  {Ngeow}, {Nordin}, {Nugent}, {Ofek}, {Patterson}, {Penprase}, {Porter},
  {Rauch}, {Rebbapragada}, {Reiley}, {Rigault}, {Rodriguez}, {van Roestel},
  {Rusholme}, {van Santen}, {Schulze}, {Shupe}, {Singer}, {Soumagnac}, {Stein},
  {Surace}, {Sollerman}, {Szkody}, {Taddia}, {Terek}, {Van Sistine}, {van
  Velzen}, {Vestrand}, {Walters}, {Ward}, {Ye}, {Yu}, {Yan}, \&
  {Zolkower}}]{bellm2019}
{Bellm}, E.~C., {Kulkarni}, S.~R., {Graham}, M.~J., {et~al.}
  2019{\natexlab{a}}, \pasp, 131, 018002, \dodoi{10.1088/1538-3873/aaecbe}

\bibitem[{{Bellm} {et~al.}(2019{\natexlab{b}}){Bellm}, {Kulkarni}, {Barlow},
  {Feindt}, {Graham}, {Goobar}, {Kupfer}, {Ngeow}, {Nugent}, {Ofek}, {Prince},
  {Riddle}, {Walters}, \& {Ye}}]{ztf_northernskysurvey_bellm}
{Bellm}, E.~C., {Kulkarni}, S.~R., {Barlow}, T., {et~al.} 2019{\natexlab{b}},
  \pasp, 131, 068003, \dodoi{10.1088/1538-3873/ab0c2a}

\bibitem[{{Belloni} {et~al.}(2020){Belloni}, {Schreiber}, {Pala},
  {G{\"a}nsicke}, {Zorotovic}, \& {Rodrigues}}]{2020belloni}
{Belloni}, D., {Schreiber}, M.~R., {Pala}, A.~F., {et~al.} 2020, \mnras, 491,
  5717, \dodoi{10.1093/mnras/stz3413}

\bibitem[{{Bernardini} {et~al.}(2017){Bernardini}, {de Martino}, {Mukai},
  {Russell}, {Falanga}, {Masetti}, {Ferrigno}, \& {Israel}}]{2017bernardini}
{Bernardini}, F., {de Martino}, D., {Mukai}, K., {et~al.} 2017, \mnras, 470,
  4815, \dodoi{10.1093/mnras/stx1494}

\bibitem[{{Bilgi}(2019)}]{wasp}
{Bilgi}, P. 2019, PhD thesis, California Institute of Technology

\bibitem[{{Boller} {et~al.}(2016){Boller}, {Freyberg}, {Tr{\"u}mper}, {Haberl},
  {Voges}, \& {Nandra}}]{2016rosat3}
{Boller}, T., {Freyberg}, M.~J., {Tr{\"u}mper}, J., {et~al.} 2016, \aap, 588,
  A103, \dodoi{10.1051/0004-6361/201525648}

\bibitem[{{Breytenbach} {et~al.}(2019){Breytenbach}, {Buckley}, {Hakala},
  {Thorstensen}, {Kniazev}, {Motsoaledi}, {Woudt}, {Potter}, {Lipunov},
  {Gorbovskoy}, {Balanutsa}, \& {Tyurina}}]{2019ecl_model}
{Breytenbach}, H., {Buckley}, D.~A.~H., {Hakala}, P., {et~al.} 2019, \mnras,
  484, 3831, \dodoi{10.1093/mnras/stz056}

\bibitem[{{Chanan} {et~al.}(1976){Chanan}, {Middleditch}, \&
  {Nelson}}]{1976chanan}
{Chanan}, G.~A., {Middleditch}, J., \& {Nelson}, J.~E. 1976, \apj, 208, 512,
  \dodoi{10.1086/154633}

\bibitem[{{Cropper}(1990)}]{1990cropper}
{Cropper}, M. 1990, \ssr, 54, 195, \dodoi{10.1007/BF00177799}

\bibitem[{{Dekany} {et~al.}(2020){Dekany}, {Smith}, {Riddle}, {Feeney},
  {Porter}, {Hale}, {Zolkower}, {Belicki}, {Kaye}, {Henning}, {Walters},
  {Cromer}, {Delacroix}, {Rodriguez}, {Reiley}, {Mao}, {Hover}, {Murphy},
  {Burruss}, {Baker}, {Kowalski}, {Reif}, {Mueller}, {Bellm}, {Graham}, \&
  {Kulkarni}}]{dekanyztf}
{Dekany}, R., {Smith}, R.~M., {Riddle}, R., {et~al.} 2020, \pasp, 132, 038001,
  \dodoi{10.1088/1538-3873/ab4ca2}

\bibitem[{{Dey} {et~al.}(2019){Dey}, {Schlegel}, {Lang}, {Blum}, {Burleigh},
  {Fan}, {Findlay}, {Finkbeiner}, {Herrera}, {Juneau}, {Landriau}, {Levi},
  {McGreer}, {Meisner}, {Myers}, {Moustakas}, {Nugent}, {Patej}, {Schlafly},
  {Walker}, {Valdes}, {Weaver}, {Y{\`e}che}, {Zou}, {Zhou}, {Abareshi},
  {Abbott}, {Abolfathi}, {Aguilera}, {Alam}, {Allen}, {Alvarez}, {Annis},
  {Ansarinejad}, {Aubert}, {Beechert}, {Bell}, {BenZvi}, {Beutler}, {Bielby},
  {Bolton}, {Brice{\~n}o}, {Buckley-Geer}, {Butler}, {Calamida}, {Carlberg},
  {Carter}, {Casas}, {Castander}, {Choi}, {Comparat}, {Cukanovaite}, {Delubac},
  {DeVries}, {Dey}, {Dhungana}, {Dickinson}, {Ding}, {Donaldson}, {Duan},
  {Duckworth}, {Eftekharzadeh}, {Eisenstein}, {Etourneau}, {Fagrelius},
  {Farihi}, {Fitzpatrick}, {Font-Ribera}, {Fulmer}, {G{\"a}nsicke},
  {Gaztanaga}, {George}, {Gerdes}, {Gontcho}, {Gorgoni}, {Green}, {Guy},
  {Harmer}, {Hernandez}, {Honscheid}, {Huang}, {James}, {Jannuzi}, {Jiang},
  {Joyce}, {Karcher}, {Karkar}, {Kehoe}, {Kneib}, {Kueter-Young}, {Lan},
  {Lauer}, {Le Guillou}, {Le Van Suu}, {Lee}, {Lesser}, {Perreault Levasseur},
  {Li}, {Mann}, {Marshall}, {Mart{\'\i}nez-V{\'a}zquez}, {Martini}, {du Mas des
  Bourboux}, {McManus}, {Meier}, {M{\'e}nard}, {Metcalfe},
  {Mu{\~n}oz-Guti{\'e}rrez}, {Najita}, {Napier}, {Narayan}, {Newman}, {Nie},
  {Nord}, {Norman}, {Olsen}, {Paat}, {Palanque-Delabrouille}, {Peng},
  {Poppett}, {Poremba}, {Prakash}, {Rabinowitz}, {Raichoor}, {Rezaie},
  {Robertson}, {Roe}, {Ross}, {Ross}, {Rudnick}, {Safonova}, {Saha},
  {S{\'a}nchez}, {Savary}, {Schweiker}, {Scott}, {Seo}, {Shan}, {Silva},
  {Slepian}, {Soto}, {Sprayberry}, {Staten}, {Stillman}, {Stupak}, {Summers},
  {Sien Tie}, {Tirado}, {Vargas-Maga{\~n}a}, {Vivas}, {Wechsler}, {Williams},
  {Yang}, {Yang}, {Yapici}, {Zaritsky}, {Zenteno}, {Zhang}, {Zhang}, {Zhou}, \&
  {Zhou}}]{2019desi_surveys}
{Dey}, A., {Schlegel}, D.~J., {Lang}, D., {et~al.} 2019, \aj, 157, 168,
  \dodoi{10.3847/1538-3881/ab089d}

\bibitem[{{Eggleton}(1983)}]{eggleton83}
{Eggleton}, P.~P. 1983, \apj, 268, 368, \dodoi{10.1086/160960}

\bibitem[{{Eker}(1992)}]{1992bydra_rscvn}
{Eker}, Z. 1992, \apjs, 79, 481, \dodoi{10.1086/191658}

\bibitem[{{Ferrario} {et~al.}(1993){Ferrario}, {Bailey}, \&
  {Wickramasinghe}}]{1993ferrario_stlmi}
{Ferrario}, L., {Bailey}, J., \& {Wickramasinghe}, D.~T. 1993, \mnras, 262,
  285, \dodoi{10.1093/mnras/262.2.285}

\bibitem[{{Ferrario} {et~al.}(2015){Ferrario}, {de Martino}, \&
  {G{\"a}nsicke}}]{2015ferrario}
{Ferrario}, L., {de Martino}, D., \& {G{\"a}nsicke}, B.~T. 2015, \ssr, 191,
  111, \dodoi{10.1007/s11214-015-0152-0}

\bibitem[{{Graham} {et~al.}(2019){Graham}, {Kulkarni}, {Bellm}, {Adams},
  {Barbarino}, {Blagorodnova}, {Bodewits}, {Bolin}, {Brady}, {Cenko}, {Chang},
  {Coughlin}, {De}, {Eadie}, {Farnham}, {Feindt}, {Franckowiak}, {Fremling},
  {Gezari}, {Ghosh}, {Goldstein}, {Golkhou}, {Goobar}, {Ho}, {Huppenkothen},
  {Ivezi{\'c}}, {Jones}, {Juric}, {Kaplan}, {Kasliwal}, {Kelley}, {Kupfer},
  {Lee}, {Lin}, {Lunnan}, {Mahabal}, {Miller}, {Ngeow}, {Nugent}, {Ofek},
  {Prince}, {Rauch}, {van Roestel}, {Schulze}, {Singer}, {Sollerman}, {Taddia},
  {Yan}, {Ye}, {Yu}, {Barlow}, {Bauer}, {Beck}, {Belicki}, {Biswas}, {Brinnel},
  {Brooke}, {Bue}, {Bulla}, {Burruss}, {Connolly}, {Cromer}, {Cunningham},
  {Dekany}, {Delacroix}, {Desai}, {Duev}, {Feeney}, {Flynn}, {Frederick},
  {Gal-Yam}, {Giomi}, {Groom}, {Hacopians}, {Hale}, {Helou}, {Henning},
  {Hover}, {Hillenbrand}, {Howell}, {Hung}, {Imel}, {Ip}, {Jackson}, {Kaspi},
  {Kaye}, {Kowalski}, {Kramer}, {Kuhn}, {Landry}, {Laher}, {Mao}, {Masci},
  {Monkewitz}, {Murphy}, {Nordin}, {Patterson}, {Penprase}, {Porter},
  {Rebbapragada}, {Reiley}, {Riddle}, {Rigault}, {Rodriguez}, {Rusholme}, {van
  Santen}, {Shupe}, {Smith}, {Soumagnac}, {Stein}, {Surace}, {Szkody}, {Terek},
  {Van Sistine}, {van Velzen}, {Vestrand}, {Walters}, {Ward}, {Zhang}, \&
  {Zolkower}}]{graham2019}
{Graham}, M.~J., {Kulkarni}, S.~R., {Bellm}, E.~C., {et~al.} 2019, \pasp, 131,
  078001, \dodoi{10.1088/1538-3873/ab006c}

\bibitem[{{Greiner} \& {Richter}(2015)}]{2015greiner}
{Greiner}, J., \& {Richter}, G.~A. 2015, \aap, 575, A42,
  \dodoi{10.1051/0004-6361/201322844}

\bibitem[{{Hailey} {et~al.}(2016){Hailey}, {Mori}, {Perez}, {Canipe}, {Hong},
  {Tomsick}, {Boggs}, {Christensen}, {Craig}, {Fornasini}, {Grindlay},
  {Harrison}, {Nynka}, {Rahoui}, {Stern}, {Zhang}, \& {Zhang}}]{2016hailey}
{Hailey}, C.~J., {Mori}, K., {Perez}, K., {et~al.} 2016, \apj, 826, 160,
  \dodoi{10.3847/0004-637X/826/2/160}

\bibitem[{{Halpern} {et~al.}(2018){Halpern}, {Thorstensen}, {Cho}, {Collver},
  {Motsoaledi}, {Breytenbach}, {Buckley}, \& {Woudt}}]{2018halpern}
{Halpern}, J.~P., {Thorstensen}, J.~R., {Cho}, P., {et~al.} 2018, \aj, 155,
  247, \dodoi{10.3847/1538-3881/aabfd0}

\bibitem[{{Hameury} \& {Lasota}(2017)}]{2017hameury_dn_ip}
{Hameury}, J.~M., \& {Lasota}, J.~P. 2017, \aap, 602, A102,
  \dodoi{10.1051/0004-6361/201730760}

\bibitem[{{Harding} {et~al.}(2016){Harding}, {Hallinan}, {Milburn}, {Gardner},
  {Konidaris}, {Singh}, {Shao}, {Sandhu}, {Kyne}, \& {Schlichting}}]{chimera}
{Harding}, L.~K., {Hallinan}, G., {Milburn}, J., {et~al.} 2016, \mnras, 457,
  3036, \dodoi{10.1093/mnras/stw094}

\bibitem[{{Harrison} \& {Campbell}(2015)}]{2015wisepolars}
{Harrison}, T.~E., \& {Campbell}, R.~K. 2015, \apjs, 219, 32,
  \dodoi{10.1088/0067-0049/219/2/32}

\bibitem[{{Harrop-Allin} {et~al.}(1999){Harrop-Allin}, {Cropper}, {Hakala},
  {Hellier}, \& {Ramseyer}}]{1999huaquarii}
{Harrop-Allin}, M.~K., {Cropper}, M., {Hakala}, P.~J., {Hellier}, C., \&
  {Ramseyer}, T. 1999, \mnras, 308, 807,
  \dodoi{10.1046/j.1365-8711.1999.02780.x}

\bibitem[{{Hellier}(2001)}]{hellierbook}
{Hellier}, C. 2001, {Cataclysmic Variable Stars}

\bibitem[{{Jackim} {et~al.}(2020){Jackim}, {Szkody}, {Hazelton}, \&
  {Benson}}]{opencvcatalog}
{Jackim}, R., {Szkody}, P., {Hazelton}, B., \& {Benson}, N.~C. 2020, Research
  Notes of the American Astronomical Society, 4, 219,
  \dodoi{10.3847/2515-5172/abd104}

\bibitem[{{Knigge} {et~al.}(2011){Knigge}, {Baraffe}, \&
  {Patterson}}]{2011knigge}
{Knigge}, C., {Baraffe}, I., \& {Patterson}, J. 2011, \apjs, 194, 28,
  \dodoi{10.1088/0067-0049/194/2/28}

\bibitem[{{Kolbin} {et~al.}(2022){Kolbin}, {Borisov}, {Serebriakova},
  {Shimansky}, {Katysheva}, {Gabdeev}, \& {Shugarov}}]{2022bstri}
{Kolbin}, A.~I., {Borisov}, N.~V., {Serebriakova}, N.~A., {et~al.} 2022,
  \mnras, 511, 20, \dodoi{10.1093/mnras/stab3676}

\bibitem[{{Kotze} {et~al.}(2015){Kotze}, {Potter}, \& {McBride}}]{2015kotze}
{Kotze}, E.~J., {Potter}, S.~B., \& {McBride}, V.~A. 2015, \aap, 579, A77,
  \dodoi{10.1051/0004-6361/201526381}

\bibitem[{{Kupfer} {et~al.}(2021){Kupfer}, {Prince}, {van Roestel}, {Bellm},
  {Bildsten}, {Coughlin}, {Drake}, {Graham}, {Klein}, {Kulkarni}, {Masci},
  {Walters}, {Andreoni}, {Biswas}, {Bradshaw}, {Duev}, {Dekany}, {Guidry},
  {Hermes}, {Laher}, \& {Riddle}}]{kupfer_ztf}
{Kupfer}, T., {Prince}, T.~A., {van Roestel}, J., {et~al.} 2021, \mnras, 505,
  1254, \dodoi{10.1093/mnras/stab1344}

\bibitem[{{Littlefield} {et~al.}(2018){Littlefield}, {Garnavich}, {Hoyt}, \&
  {Kennedy}}]{littlefield2018}
{Littlefield}, C., {Garnavich}, P., {Hoyt}, T.~J., \& {Kennedy}, M. 2018, \aj,
  155, 18, \dodoi{10.3847/1538-3881/aa9750}

\bibitem[{{Marsh}(2005)}]{doptomography}
{Marsh}, T.~R. 2005, \apss, 296, 403, \dodoi{10.1007/s10509-005-4859-3}

\bibitem[{{Masci} {et~al.}(2019){Masci}, {Laher}, {Rusholme}, {Shupe}, {Groom},
  {Surace}, {Jackson}, {Monkewitz}, {Beck}, {Flynn}, {Terek}, {Landry},
  {Hacopians}, {Desai}, {Howell}, {Brooke}, {Imel}, {Wachter}, {Ye}, {Lin},
  {Cenko}, {Cunningham}, {Rebbapragada}, {Bue}, {Miller}, {Mahabal}, {Bellm},
  {Patterson}, {Juri{\'c}}, {Golkhou}, {Ofek}, {Walters}, {Graham}, {Kasliwal},
  {Dekany}, {Kupfer}, {Burdge}, {Cannella}, {Barlow}, {Van Sistine}, {Giomi},
  {Fremling}, {Blagorodnova}, {Levitan}, {Riddle}, {Smith}, {Helou}, {Prince},
  \& {Kulkarni}}]{masci_ztf}
{Masci}, F.~J., {Laher}, R.~R., {Rusholme}, B., {et~al.} 2019, \pasp, 131,
  018003, \dodoi{10.1088/1538-3873/aae8ac}

\bibitem[{{Mason} {et~al.}(2019){Mason}, {Wells}, {Motsoaledi}, {Szkody}, \&
  {Gonzalez}}]{2019lowfieldpolar}
{Mason}, P.~A., {Wells}, N.~K., {Motsoaledi}, M., {Szkody}, P., \& {Gonzalez},
  E. 2019, \mnras, 488, 2881, \dodoi{10.1093/mnras/stz1863}

\bibitem[{{Motch} {et~al.}(1996){Motch}, {Haberl}, {Guillout}, {Pakull},
  {Reinsch}, \& {Krautter}}]{1996motch}
{Motch}, C., {Haberl}, F., {Guillout}, P., {et~al.} 1996, \aap, 307, 459

\bibitem[{{Mukai}(2017)}]{2017mukai}
{Mukai}, K. 2017, \pasp, 129, 062001, \dodoi{10.1088/1538-3873/aa6736}

\bibitem[{{O'Donoghue} {et~al.}(2006){O'Donoghue}, {Buckley}, {Balona},
  {Bester}, {Botha}, {Brink}, {Carter}, {Charles}, {Christians}, {Ebrahim},
  {Emmerich}, {Esterhuyse}, {Evans}, {Fourie}, {Fourie}, {Gajjar}, {Gordon},
  {Gumede}, {de Kock}, {Koeslag}, {Koorts}, {Kriel}, {Marang}, {Meiring},
  {Menzies}, {Menzies}, {Metcalfe}, {Meyer}, {Nel}, {O'Connor}, {Osman}, {Du
  Plessis}, {Rall}, {Riddick}, {Romero-Colmenero}, {Potter}, {Sass},
  {Schalekamp}, {Sessions}, {Siyengo}, {Sopela}, {Steyn}, {Stoffels},
  {Scholtz}, {Swart}, {Swat}, {Swiegers}, {Tiheli}, {Vaisanen}, {Whittaker}, \&
  {van Wyk}}]{2006salticam}
{O'Donoghue}, D., {Buckley}, D.~A.~H., {Balona}, L.~A., {et~al.} 2006, \mnras,
  372, 151, \dodoi{10.1111/j.1365-2966.2006.10834.x}

\bibitem[{{Oke} \& {Gunn}(1982)}]{dbsp}
{Oke}, J.~B., \& {Gunn}, J.~E. 1982, \pasp, 94, 586, \dodoi{10.1086/131027}

\bibitem[{{Oke} {et~al.}(1995){Oke}, {Cohen}, {Carr}, {Cromer}, {Dingizian},
  {Harris}, {Labrecque}, {Lucinio}, {Schaal}, {Epps}, \& {Miller}}]{lris}
{Oke}, J.~B., {Cohen}, J.~G., {Carr}, M., {et~al.} 1995, \pasp, 107, 375,
  \dodoi{10.1086/133562}

\bibitem[{{Oliveira} {et~al.}(2020){Oliveira}, {Rodrigues}, {Martins},
  {Palhares}, {Silva}, {Lima}, \& {Jablonski}}]{2020oliveiraII}
{Oliveira}, A.~S., {Rodrigues}, C.~V., {Martins}, M., {et~al.} 2020, \aj, 159,
  114, \dodoi{10.3847/1538-3881/ab6ded}

\bibitem[{{Pala} {et~al.}(2020){Pala}, {G{\"a}nsicke}, {Breedt}, {Knigge},
  {Hermes}, {Gentile Fusillo}, {Hollands}, {Naylor}, {Pelisoli}, {Schreiber},
  {Toonen}, {Aungwerojwit}, {Cukanovaite}, {Dennihy}, {Manser}, {Pretorius},
  {Scaringi}, \& {Toloza}}]{pala2020}
{Pala}, A.~F., {G{\"a}nsicke}, B.~T., {Breedt}, E., {et~al.} 2020, \mnras, 494,
  3799, \dodoi{10.1093/mnras/staa764}

\bibitem[{{Pala} {et~al.}(2022){Pala}, {G{\"a}nsicke}, {Belloni}, {Parsons},
  {Marsh}, {Schreiber}, {Breedt}, {Knigge}, {Sion}, {Szkody}, {Townsley},
  {Bildsten}, {Boyd}, {Cook}, {De Martino}, {Godon}, {Kafka}, {Kouprianov},
  {Long}, {Monard}, {Myers}, {Nelson}, {Nogami}, {Oksanen}, {Pickard},
  {Poyner}, {Reichart}, {Rodriguez Perez}, {Shears}, {Stubbings}, \&
  {Toloza}}]{pala2022}
{Pala}, A.~F., {G{\"a}nsicke}, B.~T., {Belloni}, D., {et~al.} 2022, \mnras,
  510, 6110, \dodoi{10.1093/mnras/stab3449}

\bibitem[{{Predehl} {et~al.}(2021){Predehl}, {Andritschke}, {Arefiev},
  {Babyshkin}, {Batanov}, {Becker}, {B{\"o}hringer}, {Bogomolov}, {Boller},
  {Borm}, {Bornemann}, {Br{\"a}uninger}, {Br{\"u}ggen}, {Brunner}, {Brusa},
  {Bulbul}, {Buntov}, {Burwitz}, {Burkert}, {Clerc}, {Churazov}, {Coutinho},
  {Dauser}, {Dennerl}, {Doroshenko}, {Eder}, {Emberger}, {Eraerds},
  {Finoguenov}, {Freyberg}, {Friedrich}, {Friedrich}, {F{\"u}rmetz},
  {Georgakakis}, {Gilfanov}, {Granato}, {Grossberger}, {Gueguen}, {Gureev},
  {Haberl}, {H{\"a}lker}, {Hartner}, {Hasinger}, {Huber}, {Ji}, {Kienlin},
  {Kink}, {Korotkov}, {Kreykenbohm}, {Lamer}, {Lomakin}, {Lapshov}, {Liu},
  {Maitra}, {Meidinger}, {Menz}, {Merloni}, {Mernik}, {Mican}, {Mohr},
  {M{\"u}ller}, {Nandra}, {Nazarov}, {Pacaud}, {Pavlinsky}, {Perinati},
  {Pfeffermann}, {Pietschner}, {Ramos-Ceja}, {Rau}, {Reiffers}, {Reiprich},
  {Robrade}, {Salvato}, {Sanders}, {Santangelo}, {Sasaki}, {Scheuerle},
  {Schmid}, {Schmitt}, {Schwope}, {Shirshakov}, {Steinmetz}, {Stewart},
  {Str{\"u}der}, {Sunyaev}, {Tenzer}, {Tiedemann}, {Tr{\"u}mper}, {Voron},
  {Weber}, {Wilms}, \& {Yaroshenko}}]{2021erosita}
{Predehl}, P., {Andritschke}, R., {Arefiev}, V., {et~al.} 2021, \aap, 647, A1,
  \dodoi{10.1051/0004-6361/202039313}

\bibitem[{{Prsa} \& {Zwitter}(2005)}]{2005phoebe1}
{Prsa}, A., \& {Zwitter}, T. 2005, \apj, 628, 426, \dodoi{10.1086/430591}

\bibitem[{{Prsa} {et~al.}(2016){Prsa}, {Conroy}, {Horvat}, {Pablo}, {Kochoska},
  {Bloemen}, {Giammarco}, {Hambleton}, \& {Degroote}}]{2016phoebe2}
{Prsa}, A., {Conroy}, K.~E., {Horvat}, M., {et~al.} 2016, \apjs, 227, 29,
  \dodoi{10.3847/1538-4365/227/2/29}

\bibitem[{{Rosswog} \& {Br{\"u}ggen}(2007)}]{2007rosswog}
{Rosswog}, S., \& {Br{\"u}ggen}, M. 2007, {Introduction to High-Energy
  Astrophysics}

\bibitem[{{Salvato} {et~al.}(2021){Salvato}, {Wolf}, {Dwelly}, {Georgakakis},
  {Brusa}, {Merloni}, {Liu}, {Toba}, {Nandra}, {Lamer}, {Buchner}, {Schneider},
  {Freund}, {Rau}, {Schwope}, {Nishizawa}, {Klein}, {Arcodia}, {Comparat},
  {Musiimenta}, {Nagao}, {Brunner}, {Malyali}, {Finoguenov}, {Anderson},
  {Shen}, {Ibarra-Mendel}, {Trump}, {Brandt}, {Urry}, {Rivera}, {Krumpe},
  {Urrutia}, {Miyaji}, {Ichikawa}, {Schneider}, {Fresco}, {Wilms}, {Boller},
  {Haase}, {Brownstein}, {Lane}, {Bizyaev}, \& {Nitschelm}}]{salvato2021}
{Salvato}, M., {Wolf}, J., {Dwelly}, T., {et~al.} 2021, arXiv e-prints,
  arXiv:2106.14520.
\newblock \doarXiv{2106.14520}

\bibitem[{{Schmidt} {et~al.}(2005){Schmidt}, {Szkody}, {Homer}, {Smith},
  {Chen}, {Henden}, {Solheim}, {Wolfe}, \& {Greimel}}]{schmidt2005}
{Schmidt}, G.~D., {Szkody}, P., {Homer}, L., {et~al.} 2005, \apj, 620, 422,
  \dodoi{10.1086/426807}

\bibitem[{{Schreiber} {et~al.}(2021){Schreiber}, {Belloni}, {G{\"a}nsicke},
  {Parsons}, \& {Zorotovic}}]{schreiber2021}
{Schreiber}, M.~R., {Belloni}, D., {G{\"a}nsicke}, B.~T., {Parsons}, S.~G., \&
  {Zorotovic}, M. 2021, Nature Astronomy, 5, 648,
  \dodoi{10.1038/s41550-021-01346-8}

\bibitem[{{Schwope} {et~al.}(2021){Schwope}, {Buckley}, {Malyali}, {Potter},
  {K{\"o}nig}, {Arcodia}, {Gromadzki}, \& {Rau}}]{schwope2021}
{Schwope}, A., {Buckley}, D. A.~H., {Malyali}, A., {et~al.} 2021, arXiv
  e-prints, arXiv:2106.14540.
\newblock \doarXiv{2106.14540}

\bibitem[{{Schwope} {et~al.}(2011){Schwope}, {Horne}, {Steeghs}, \&
  {Still}}]{2011schwope}
{Schwope}, A.~D., {Horne}, K., {Steeghs}, D., \& {Still}, M. 2011, \aap, 531,
  A34, \dodoi{10.1051/0004-6361/201016373}

\bibitem[{{Sheinis} {et~al.}(2002){Sheinis}, {Bolte}, {Epps}, {Kibrick},
  {Miller}, {Radovan}, {Bigelow}, \& {Sutin}}]{esi}
{Sheinis}, A.~I., {Bolte}, M., {Epps}, H.~W., {et~al.} 2002, \pasp, 114, 851,
  \dodoi{10.1086/341706}

\bibitem[{{Silber}(1992)}]{1992silber}
{Silber}, A.~D. 1992, PhD thesis, Massachusetts Institute of Technology

\bibitem[{{Sunyaev} {et~al.}(2021){Sunyaev}, {Arefiev}, {Babyshkin},
  {Bogomolov}, {Borisov}, {Buntov}, {Brunner}, {Burenin}, {Churazov},
  {Coutinho}, {Eder}, {Eismont}, {Freyberg}, {Gilfanov}, {Gureyev}, {Hasinger},
  {Khabibullin}, {Kolmykov}, {Komovkin}, {Krivonos}, {Lapshov}, {Levin},
  {Lomakin}, {Lutovinov}, {Medvedev}, {Merloni}, {Mernik}, {Mikhailov},
  {Molodtsov}, {Mzhelsky}, {M{\"u}ller}, {Nandra}, {Nazarov}, {Pavlinsky},
  {Poghodin}, {Predehl}, {Robrade}, {Sazonov}, {Scheuerle}, {Shirshakov},
  {Tkachenko}, \& {Voron}}]{2021sunyaev}
{Sunyaev}, R., {Arefiev}, V., {Babyshkin}, V., {et~al.} 2021, \aap, 656, A132,
  \dodoi{10.1051/0004-6361/202141179}

\bibitem[{{Szkody} {et~al.}(2021){Szkody}, {Olde Loohuis}, {Koplitz}, {van
  Roestel}, {Dicenzo}, {Ho}, {Hillenbrand}, {Bellm}, {Dekany}, {Drake}, {Duev},
  {Graham}, {Kasliwal}, {Mahabal}, {Masci}, {Neill}, {Riddle}, {Rusholme},
  {Sollerman}, \& {Walters}}]{ztf_cv_yr2}
{Szkody}, P., {Olde Loohuis}, C., {Koplitz}, B., {et~al.} 2021, \aj, 162, 94,
  \dodoi{10.3847/1538-3881/ac0efb}

\bibitem[{{Truemper}(1982)}]{1982rosat1}
{Truemper}, J. 1982, Advances in Space Research, 2, 241,
  \dodoi{10.1016/0273-1177(82)90070-9}

\bibitem[{{VanderPlas}(2016)}]{gatspy}
{VanderPlas}, J. 2016, {gatspy: General tools for Astronomical Time Series in
  Python}.
\newblock \doeprint{1610.007}

\bibitem[{{VanderPlas}(2018)}]{2018vanderplas}
{VanderPlas}, J.~T. 2018, \apjs, 236, 16, \dodoi{10.3847/1538-4365/aab766}

\bibitem[{{Voges} {et~al.}(1999){Voges}, {Aschenbach}, {Boller},
  {Br{\"a}uninger}, {Briel}, {Burkert}, {Dennerl}, {Englhauser}, {Gruber},
  {Haberl}, {Hartner}, {Hasinger}, {K{\"u}rster}, {Pfeffermann}, {Pietsch},
  {Predehl}, {Rosso}, {Schmitt}, {Tr{\"u}mper}, \& {Zimmermann}}]{1999rosat2}
{Voges}, W., {Aschenbach}, B., {Boller}, T., {et~al.} 1999, \aap, 349, 389.
\newblock \doarXiv{astro-ph/9909315}

\bibitem[{{Warner}(1995)}]{warner95}
{Warner}, B. 1995, {Cataclysmic variable stars}, Vol.~28

\bibitem[{{Warwick} {et~al.}(2012){Warwick}, {Saxton}, \&
  {Read}}]{2012galex_misclassification}
{Warwick}, R.~S., {Saxton}, R.~D., \& {Read}, A.~M. 2012, \aap, 548, A99,
  \dodoi{10.1051/0004-6361/201118642}

\bibitem[{{Wickramasinghe} \& {Ferrario}(2000)}]{2000oldmagrev}
{Wickramasinghe}, D.~T., \& {Ferrario}, L. 2000, \pasp, 112, 873,
  \dodoi{10.1086/316593}

\bibitem[{{Zhilkin} {et~al.}(2019){Zhilkin}, {Sobolev}, {Bisikalo}, \&
  {Gabdeev}}]{2019ecl_model2}
{Zhilkin}, A.~G., {Sobolev}, A.~V., {Bisikalo}, D.~V., \& {Gabdeev}, M.~M.
  2019, Astronomy Reports, 63, 751, \dodoi{10.1134/S1063772919090087}

\end{thebibliography}
\bibliographystyle{aasjournal}

\end{document}